\documentclass[aps,prb,reprint,superscriptaddress,groupedaddress,amsmath,amssymb]{revtex4-2}
\usepackage[dvipdfmx]{graphicx}
\usepackage{dcolumn}
\usepackage{bm}
\usepackage{braket}
\usepackage{hyperref}
\usepackage{amsmath,amssymb}
\usepackage{color}
\usepackage{comment}
\hypersetup{
colorlinks = true, 
urlcolor = blue, 
linkcolor = blue, 
citecolor = blue
}

\begin{document}

\title{Doublon-holon pairing state in photodoped Mott insulators}
\author{
Ryota Ueda$^1$, 
Madhumita Sarkar$^{2,3}$,
Zala Lenar\v{c}i\v{c}$^2$,
Denis Gole\v{z}$^{2,4}$,
Kazuhiko Kuroki$^1$, 
Tatsuya Kaneko$^1$
}
\affiliation{
$^1$Department of Physics, The University of Osaka, Toyonaka, Osaka 560-0043, Japan \\
$^2$Jo\v{z}ef Stefan Institute, Jamova 39, SI-1000 Ljubljana, Slovenia \\
$^3$Department of Physics and Astronomy, University of Exeter, Stocker Road, Exeter EX4 4QL, United Kingdom\\
$^4$Faculty of Mathematics and Physics, University of Ljubljana, Jadranska 19, 1000 Ljubljana, Slovenia
}
\date{\today}
\begin{abstract}
We demonstrate the existence of an unconventional pairing state in photodoped Mott insulators on ladder and quasi-two-dimensional geometries, characterized by quasi-long-range doublon-holon correlations that signal Mott exciton condensation. 
The doublon-holon pairing exhibits correlations of $d$-wave-like symmetry, reminiscent of superconducting pairing in chemically doped Mott insulators.
By constructing the phase diagram, using density matrix renormalization group, we reveal that the doublon-holon pairing state in the photodoped ladder emerges between the spin-singlet, charge-density-wave, and $\eta$-pairing phases.   
Our study suggests that the interplay of charge, spin, and $\eta$-spin degrees of freedom can give rise to exotic quantum many-body states in photodoped Mott insulators.
\end{abstract}

\maketitle

\begin{figure}[!b]
    \begin{center}
        \includegraphics[width=\columnwidth]{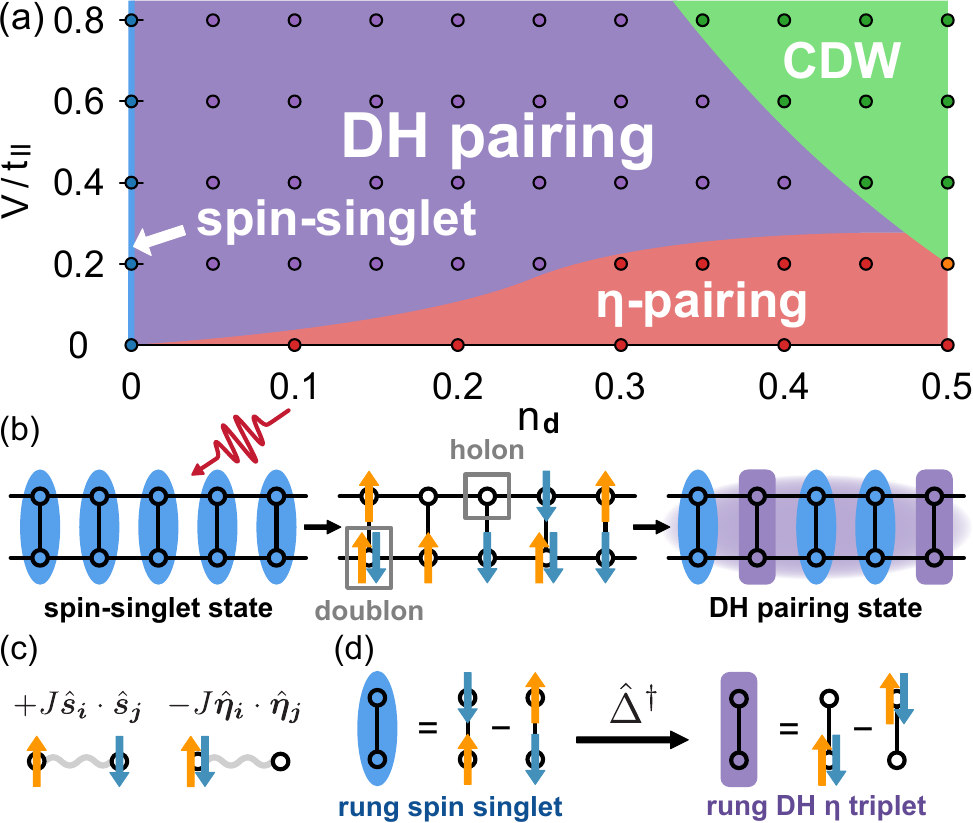}
        \caption{(a) Phase diagram of the photodoped ladder as a function of the doublon density $n_d$ and the intersite Coulomb interaction $V$ with the blue, red, green, and purple points representing the predominance of the spin, $\eta$-pairing, CDW, and doublon-holon (DH) pairing correlations, respectively. 
        At the orange point ($n_d = 0.5$ and $V = 0.2t_{\parallel}$), the $\eta$-pairing and CDW states are degenerate. 
        Parameters: $t_{\perp}=t_{\parallel}$, $J_{\perp}=J_{\parallel} = 0.4 t_{\parallel}$, and $V=V_{\parallel} = V_{\perp}$. 
        (b) DH pairing state in the photodoped ladder with doublons and holons generated by light. 
        (c) Spin interaction that acts between up and down spins, and $\eta$-spin interaction that acts between a doublon and a holon. 
        (d) Operator that maps a spin singlet on nearest-neighbor sites to a DH $\eta$ triplet. 
        }
        \label{phase}
    \end{center}
\end{figure}

\textit{Introduction}. External field driving is a useful tool for generating and controlling intriguing out-of-equilibrium phenomena in correlated electron systems~\cite{yonemitsu2008, claudio2016, basov2017, ishihara2019, torre2021,murakami2023_2}.
Photoinduced phase transitions~\cite{fausti2011, mitrano2016, okazaki2018, cavalleri2018, budden2021, saha2021, koshihara2022}, Floquet engineering~\cite{mentink2015, claassen2017, oka2019, kitamura2022, takahashi2025}, and high-harmonic generation~\cite{silva2018, murakami2018, ghimire2019, imai2020, udono2022, uchida2022, ono2024, murakami2025} have been extensively studied to date.   
In semiconductors, optical excitation across the band gap generates electron and hole carriers. 
Similarly, photoexcitation in correlation-driven Mott insulators (MIs) creates two types of carriers, doublons (doubly occupied sites) and holons (empty sites) [see Fig.~\ref{phase}(b)]~\cite{diasdasilva2012,terashige2019,murakami2023_2,huang2023,tsutsui2023,bohrdt_arXiv}, resulting in a photodoped state distinct from conventional photoexcited semiconductors.
In a large-gap MI, the carriers are expected to have long lifetimes, leading to a quasisteady state where doublons and holons coexist~\cite{rosch2008, strohmaier2010, sensarma2010, eckstein2011,zala2013, lenarcic2014, lenarcic2015, mitrano2014,murakami2022, murakami2023} and possibly bind in excitons \cite{takahashi2002,tohyama2006,zala2013,lenarcic2014,lenarcic2015,mehio2023}, which are distinct from excitons in photodoped semiconductors~\cite{keldysh1986,yoshioka2011}.

Photodoping can be a trigger for creating exotic many-body states that are inaccessible in equilibrium, such as the $\eta$-pairing state, which is characterized by long-range correlation of on-site electronic pairs~\cite{yang1989, kaneko2019, peronaci2020, ejima2020, kaneko2020, tindall2020, ejima2022, ueda2024, ray2024,imai2025}. 
The photodoped states have been theoretically investigated in detail in the one-dimensional (1D) Hubbard chain~\cite{murakami2022, murakami2023} and in high-dimensional Hubbard systems~\cite{werner2019,li2020,li2023}. 
In the former case, the phase diagram comprises the spin-density-wave (SDW) phase at low doublon density $n_d$ and the charge-density-wave (CDW) and $\eta$-pairing phases from intermediate to high $n_d$~\cite{murakami2022,murakami2023}. 
Although even richer physical properties are anticipated in the two-dimensional (2D) systems because of the increased geometric degrees of freedom and the lack of spin-charge-$\eta$ separation~\cite{murakami2023}, these properties of photodoped 2D MIs are yet to be understood.

Ladder geometry can serve as a bridge between 1D and 2D systems and is more manageable for numerical approaches. 
For example, density matrix renormalization group (DMRG) calculations, which enable us to obtain numerically precise low-dimensional wave functions, have shown that chemically hole-doped ladders exhibit pair correlations corresponding to $d$-wave superconductivity (SC) in a 2D square lattice~\cite{dagotto1994,noack1994,noack1997,dolfi2015,sheikhan2020, shen2023}. 
These studies are an important stepping stone to research on $d$-wave SC in the Hubbard and $t$-$J$ models on multileg cylinders, mimicking the 2D square lattice of cuprate superconductors~\cite{hong-chen2019,chung2020,jiang2021,gong2021,shengtao2021,lu2024}.
Furthermore, the pairing mechanism in ladder and bilayer systems~\cite{dagotto1992,kuroki2002,maier2011,bohrdt2022,zhang2023,qu2024,kaneko2024,sakakibara2024,zhang2024,kakoi2024,luo2024,schlomer2024} has attracted significant attention associated with the recent discovery of high-temperature SC in bilayer nickelates~\cite{sun2023,sakakibara2024_4310,wang2024,ko2025}. 
Because a strong interchain spin coupling favors the formation of spin-singlet rungs, hole doping to the ladder system leads to the formation of interchain hole-hole pairs to minimize the disruption of spin-singlet bonds~\cite{dagotto1992}. 
Realization of such a pairing mechanism has also been recently proposed with cold atoms in optical lattices~\cite{hirthe2023}. 
From these considerations, we arrive at the crucial question: what are the equivalent correlation-driven paired states in photodoped MIs?

In this Letter, we show that photodoped MIs in a quasistationary situation exhibit a pairing state composed of doublons and holons (Fig.~\ref{phase}).  
In the ladder geometry, this pairing state is characterized by the quasi-long-range correlation of doublon-holon (DH) pairs (Mott excitons), suggesting exciton condensation.
Due to strong correlations, the state is distinct from excitonic condensates in semiconductors. 
Namely, we observe opposite phases for the rung and chain pairing, analogous to signatures of superconducting $d$-wave pairing in chemically doped ladders. 
These findings are further corroborated by results on a multileg cylinder mimicking the 2D square lattice.

\textit{Model}. To model MIs in the presence of nearest-neighbor interactions, we start with the standard extended Hubbard Hamiltonian
\begin{align}
\hat{\mathcal{H}}_{\mathrm{Hub}} =  
\sum_{\langle\bm{ij}\rangle}  \hat{T}_{\bm{ij}}
+ U \sum_{\bm{j}} \hat{n}'_{\bm{j}; \uparrow} \hat{n}'_{\bm{j}; \downarrow} 
+ \sum_{\langle\bm{ij}\rangle}  V_{\bm{ij}} \hat{n}'_{\bm{i}} \hat{n}'_{\bm{j}} , 
\label{Hamiltonian}
\end{align}
where $\hat{T}_{\bm{ij}} = - t_{\bm{ij}}\sum_{\sigma} \hat{c}^\dagger_{\bm{i}; \sigma} \hat{c}_{\bm{j}; \sigma} + {\rm H.c.}$ is the hopping term. 
$\hat{c}^\dagger_{\bm{j}; \sigma}$ ($\hat{c}_{\bm{j}; \sigma}$) is the fermion creation (annihilation) operator at site $\bm{j}$ for spin $\sigma = \uparrow,$~$\downarrow$. 
$\hat{n}_{\bm{j}; \sigma} = \hat{c}^\dagger_{\bm{j}; \sigma} \hat{c}_{\bm{j}; \sigma}$,
$\hat{n}'_{\bm{j}; \sigma} = \hat{n}_{\bm{j}; \sigma} - 1 / 2$, and
$\hat{n}'_{\bm{j}} = \hat{n}'_{\bm{j}; \uparrow} + \hat{n}'_{\bm{j}; \downarrow}$.
Summation $\braket{\bm{i}\bm{j}}$ runs over nearest-neighbor sites. 
In the two-leg ladder, $\bm{j} =(j,\alpha)$ denotes site at position $j$ on chain $\alpha$ [$=0$ (lower), $1$ (upper)], 
and we distinguish the hopping integrals $t_{\bm{ij}}=t_{\parallel}, t_{\perp}$ and the nearest-neighbor Coulomb interaction $V_{\bm{ij}}=V_{\parallel}, V_{\perp}$ along the chain and rung directions, respectively. 
In the 2D square lattice, $t_{\bm{ij}}=t_{\rm h}$ and $V_{\bm{ij}}=V$.
$U$ is the on-site Coulomb repulsion.
The lattice constant $a$ is set to one, and we use $t_\parallel$ in the two-leg ladder ($t_{\rm h}$ in the square lattice) as the unit of energy unless otherwise specified. 

In photodoped MIs, doublons and holons are generated by an optical excitation, as shown in Fig.~\ref{phase}(b). 
Their recombination rate $\Gamma_{\rm R} \propto \exp(- U/\epsilon_0)$ is exponentially suppressed with the number of excitations that are emitted in the recombination process~\cite{rosch2008, strohmaier2010, sensarma2010, eckstein2011, zala2013, lenarcic2014,mitrano2014,lenarcic2015}. 
For low densities of doublons and holons, the recombination predominantly happens via emission of magnetic processes, $\epsilon_0$ being their characteristic energy~\cite{mehio2023}. 
For higher densities, scattering with other doublons and holons ($\epsilon_0\sim t_{\bm{ij}}$) can also contribute \cite{strohmaier2010,sensarma2010}. 
Thus, when $U\gg \epsilon_0$, the system can be on the timescales that are parametrically long in $U$ considered to be in a pseudoequilibrium state where doublons and holons are conserved~\cite{murakami2022, murakami2023}.
To efficiently consider the case $U \gg \epsilon_0$, we introduce an effective model derived by the Schrieffer-Wolff transformation that conserves the number of doublons and holons~\cite{zala2013,lenarcic2014,murakami2022, sarkar2024} with the dominant (lowest order) terms:
\begin{align}\label{eff} 
\hat{\mathcal{H}} = 
&-\sum_{\langle\bm{ij}\rangle,\sigma} t_{\bm{ij}} 
\left( 
\hat{d}^\dagger_{\bm{i}; \sigma} \hat{d}_{\bm{j}; \sigma}  
- \hat{h}^\dagger_{\bm{i}; \sigma} \hat{h}_{\bm{j}; \sigma} 
+ {\rm H.c.} \right) 
\notag \\
&+\sum_{\langle\bm{ij}\rangle} J_{\bm{i}\bm{j}} \left( 
\hat{\bm{s}}_{\bm{i}} \cdot \hat{\bm{s}}_{\bm{j}} 
-\frac{1}{4}
\delta_{1,\hat{n}_{\bm{i}}} \delta_{1,\hat{n}_{\bm{j}}}
\right) 
\\
&- \sum_{\langle\bm{ij}\rangle}\left[ J_{\bm{i}\bm{j}} \left( 
\hat{\bm{\eta}}_{\bm{i}} \cdot \hat{\bm{\eta}}_{\bm{j}} 
-\frac{1}{4}\bar{\delta}_{1,\hat{n}_{\bm{i}}} \bar{\delta}_{1,\hat{n}_{\bm{j}}}
\right)  
-4V_{\bm{i}\bm{j}}\hat{\eta}^z_{\bm{i}}\hat{\eta}^z_{\bm{j}} \right].
\notag
\end{align}
The first term denotes the hopping of doublons and holons, written explicitly with doublon $\hat{d}^\dagger_{\bm{j};\sigma}=\hat{c}^\dagger_{\bm{j};\sigma}\hat{n}_{\bm{j};\bar{\sigma}}$ and holon $\hat{h}^\dagger_{\bm{j};\sigma}=\hat{c}_{\bm{j};\sigma}(1-\hat{n}_{\bm{j};\bar{\sigma}})$ creation operators \cite{zala2013}, where $\bar{\sigma}=\downarrow$~$(\uparrow)$ for $\sigma=\uparrow$~$(\downarrow)$.
The second term describes the SU$(2)$ symmetric spin-exchange interaction, which is also considered in the traditional $t$-$J$ model. 
The spin operator is defined as $\hat{\bm{s}}_{\bm{j}} = \sum_{\sigma, \sigma'} \hat{c}^\dagger_{\bm{j};\sigma} \bm{\sigma}_{\sigma\sigma'} \hat{c}_{\bm{j};\sigma'} / 2$, where $\bm{\sigma}$ is the vector of Pauli matrices. 
For the hopping $t_{\bm{i}\bm{j}}$, the coupling constant is given by $J_{\bm{i}\bm{j}}=4t^2_{\bm{i}\bm{j}}/U > 0$ (at $V_{\bm{i}\bm{j}}=0$). 
Similarly, the third term represents the interaction between doublons and holons~\cite{murakami2022, sarkar2024}. 
In the absence of $V_{\bm{i}\bm{j}}$, this term reveals the SU$(2)$ symmetry of the Hubbard model, represented by the $\eta$ spin. 
The $\eta$-spin operator $\hat{\bm{\eta}}_{\bm{j}}$ is given by $\hat{\eta}^+_{\bm{j}} = (-1)^{j+\alpha} \hat{c}^\dagger_{\bm{j}; \downarrow} \hat{c}^\dagger_{\bm{j}; \uparrow}$, $\hat{\eta}^-_{\bm{j}} = (-1)^{j+\alpha} \hat{c}_{\bm{j}; \uparrow} \hat{c}_{\bm{j}; \downarrow}$, and $\hat{\eta}^z_{\bm{j}} = \left( \hat{n}_{\bm{j}} - 1\right) / 2$~\cite{essler2005}. 
In Eq.~\eqref{eff}, $\bar{\delta}_{1,\hat{n}_{\bm{j}}}=1-\delta_{1,\hat{n}_{\bm{j}}}$, and the $\eta$ term acts only on empty ($\circ$) and doubly occupied ($\uparrow\downarrow$) sites. 
$\hat{\eta}^+_{\bm{i}}\hat{\eta}^-_{\bm{j}}$ swaps the positions of neighboring doublon and holon, while $\hat{\eta}^z_{\bm{i}}\hat{\eta}^z_{\bm{j}}$ measures the energy according to the occupancy status of sites $\bm{i}$ and $\bm{j}$~\cite{SM}. 
The $\eta$-spin interaction in Eq.~(\ref{eff}) is ferromagnetic, which preferably forms an $\eta$-spin triplet state at two sites. 
This $\eta$-spin interaction is activated by photodoping, while it is negligible in the traditional $t$-$J$ model for chemically doped systems.
As is clear from Eq.~\eqref{eff}, $V_{\bm{i}\bm{j}}$ manifestly breaks the $\eta$-SU$(2)$ symmetry and leads to an anisotropic XXZ-type Hamiltonian for the $\eta$ sector. 
Overall, $\hat{\mathcal{H}}$ is expected to describe metastable phases in photodoped MIs with a similar level of rigor as the well-known $t$-$J$ model can describe chemically doped materials.

For $U\gg t_{\boldsymbol{ij}}, J_{\boldsymbol{ij}}$, implying slow recombination of doublons and holons, photoexcited carriers have time to relax to the bottom of the upper band. 
Therefore, we can transiently describe the photodoped state as the lowest-energy state of $\hat{\mathcal{H}}$ at a fixed doublon density $n_d$ that corresponds to the excitation protocol. 
In the two-leg ladder of length $L$, the doublon density is given by $n_d=\sum_{j, \alpha} \braket{\hat{n}_{j,\alpha;\uparrow}\hat{n}_{j,\alpha;\downarrow}}/(2L)$. 
The lowest-energy state, representing the transient photodoped state, is obtained with the DMRG method~\cite{white1992, white1993, schollwock2011} within a sector with a fixed number of doublons~\cite{zigzag}.
In the two-leg ladder, the total number of sites is $N_{\rm s}=2L$.  
The numbers of up- and down-spin fermions are set to $N_\uparrow= N_\downarrow = N_{\rm s}/2$ (i.e., half filling without spin polarization), while the state is configured as a function of the doublon number.  
The given doublon number $N_d$ determines the doublon density as $n_d = N_d /N_{\rm s}$.
We employ open boundary conditions along the chain ($x$) direction. 
In the two-leg ladder, we set the chain length to $L = 160$ unless otherwise specified.
The bond dimension is set to $m = 10000$ with the maximum truncation error typically being on the order of $1.0 \times 10^{-7}$. 
For the 2D square lattice, we employ an $8 \times 4$ site cylinder with periodic boundary conditions along the $y$ direction, where the bond dimension and the maximum truncation error are $m = 15000$ and $1.0 \times 10^{-5}$, respectively.

To identify nonequilibrium phases in photodoped states, we analyze correlation functions $C(r) = \braket{\hat{O}^\dagger_{j_0 + r}\hat{O}_{j_0}}$ between operators \(\hat{O}_j\) separated by \(r\) sites.
The reference site is fixed at $j_0 = L/4 + 1$ 
with open boundary conditions.
We present the site-averaged correlation functions in the Supplemental Material~\cite{SM} and confirm that our conclusion does not rely on the definition of the correlation function. 
The dominant phase is determined by the correlation function that exhibits the slowest decay with distance $r$.

\begin{figure}[t]
    \begin{center}       
        \includegraphics[width=\columnwidth]{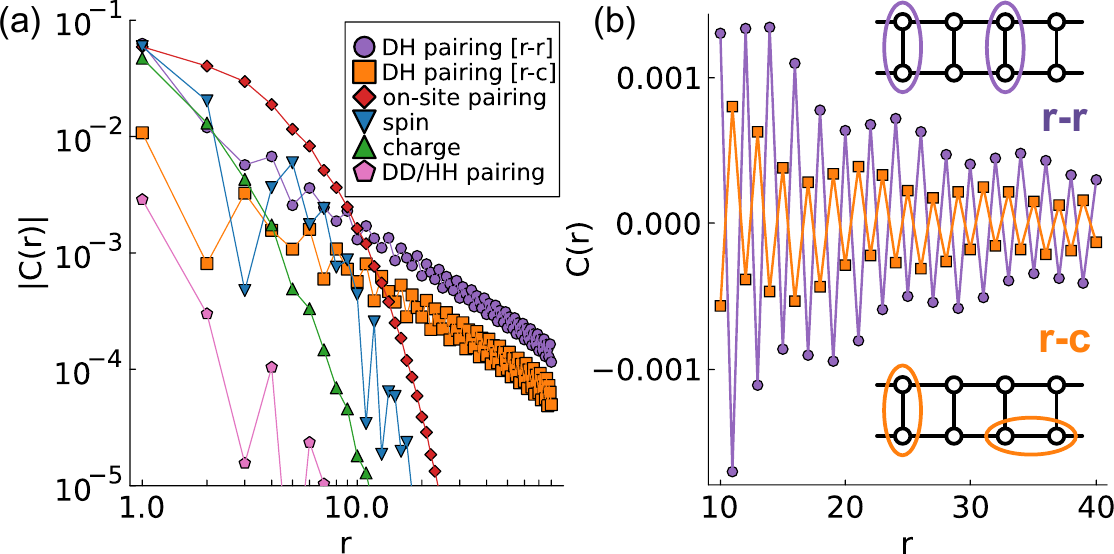}        
        \caption{(a) Log-log plot and (b) linear scale plot of the correlation functions for $n_d = 0.2$,  $t_\perp = t_\parallel$, $J_\perp = J_\parallel = 0.4t_\parallel$, and $V = V_{\perp} = V_{\parallel} = 0.2t_\parallel$. 
        r-r (purple circles) and r-c (orange squares) represent the rung-rung and rung-chain DH pairing correlations, respectively. 
        The on-site pairing correlations are multiplied by two for comparison on the same scale.} 
        \label{correlation}
    \end{center}
\end{figure}

\textit{Numerical results}.
The central result of this Letter is the finding of a DH pairing phase, characterized with the slowest power-law decay in the correlation functions $\braket{\hat{\Delta}^{\mathrm{r}\dagger}_{j_0+r}\hat{\Delta}^\mathrm{r}_{j_0}}$ and $\braket{\hat{\Delta}^{\mathrm{c}\dagger}_{j_0+r}\hat{\Delta}^\mathrm{r}_{j_0}}$, 
defined via the operators 
\begin{align}
\hat{\Delta}^\mathrm{\mathrm{r}\dagger}_{j} &= \frac{1}{2}\sum_{\alpha=0,1} \sum_{\sigma} (-1)^{\alpha}
\hat{d}^\dagger_{j,\alpha;\sigma}
\hat{h}^\dagger_{j,\bar{\alpha};\sigma}, 
\\
\hat{\Delta}^\mathrm{\mathrm{c}\dagger}_{j} &= \frac{1}{2}\sum_{\beta=0,1} \sum_{\sigma} (-1)^{\beta}
\hat{d}^\dagger_{j+\beta,0;\sigma}
\hat{h}^\dagger_{j+\bar{\beta},0;\sigma}, 
\end{align}
where $\bar{\alpha}=1$~$(0)$ for $\alpha=0$~$(1)$. 
These operators act on nearest-neighboring sites that are in the spin-singlet state and transfer it to the state with a doublon and a holon in the $\eta$-triplet state, as shown in Fig.~\ref{phase}(d), along the rung (r) and chain (c) directions, respectively. 
The definitions of the pair operators given above correspond to the formation of a local component of a Mott exciton, energetically favored by the interactions between nearest-neighbor sites (see 
Supplemental Material~\cite{SM} for details).

Figure~\ref{correlation} shows the correlation functions at $n_d = 0.2$ and $V = 0.2t_\parallel$ in the isotropic ladder ($t_{\perp}=t_{\parallel}$, $V_{\perp}=V_{\parallel}=V$). 
The log-log plot in Fig.~\ref{correlation}(a) shows that the DH pairing correlation $\braket{\hat{\Delta}^{\mathrm{r}\dagger}_{j_0+r}\hat{\Delta}^\mathrm{r}_{j_0}}$ exhibits a power-law decay, while the spin ($\hat{O}_{j} = \hat{n}_{j,\alpha;\uparrow} - \hat{n}_{j,\alpha; \downarrow}$), charge ($\hat{O}_{j} = \hat{n}_{j,\alpha} - 1$), and on-site pair ($\hat{O}_{j} = \hat{c}_{j,\alpha;\uparrow} \hat{c}_{j,\alpha; \downarrow}$) correlations~\cite{CF} decay exponentially, indicating that the DH pairing correlation is dominant. 
As shown in Fig.~\ref{correlation}(b), the rung-rung correlation $\braket{\hat{\Delta}^{\mathrm{r}\dagger}_{j_0+r}\hat{\Delta}^{\mathrm{r}}_{j_0}}$ and the rung-chain correlation $\braket{\hat{\Delta}^{\mathrm{c}\dagger}_{j_0+r}\hat{\Delta}^{\mathrm{r}}_{j_0}}$ exhibit opposite signs. 
Similar sign inversion between superconducting rung-rung and rung-chain correlators appears in chemically doped ladders~\cite{SM}, which was taken as a signature of $d$-wave pairing~\cite{noack1994}. 
This intriguing finding suggests that the DH pairing correlation in 2D systems potentially bears a close similarity to the equilibrium $d$-wave SC state in chemically doped MIs. 

In principle, a superconducting pairing state of holons could also arise in the transient nonequilibrium states of the photodoped systems, since the protocol, similar to chemical doping, creates a finite holon density. 
To compare the traditional holon-holon (HH) pairing with the DH pairing in our photodoped state, we introduce the holon-holon pairing operator 
$\hat{\Delta}_j^{{\rm HH}\dagger} = \frac{1}{\sqrt{2}} \sum_{\alpha} \hat{h}^\dagger_{j,\alpha;\downarrow} \hat{h}^\dagger_{j,\bar{\alpha};\uparrow}$, which maps the spin singlet to the $\ket{00}$ state on the rung.
Analogously, the doublon-doublon (DD) pairing operator $\hat{\Delta}_j^{\mathrm{DD}\dagger} = \frac{1}{\sqrt{2}} \sum_{\alpha} \hat{d}^\dagger_{j,\alpha;\downarrow} \hat{d}^\dagger_{j,\bar{\alpha};\uparrow}$ has a correlation function equivalent to that of holons in the photodoped case. 
Figure~\ref{correlation}(a) shows that $\braket{\hat{\Delta}^{{\rm HH}\dagger}_{j_0+r}\hat{\Delta}^{\mathrm{HH}}_{j_0}}$ decays faster and is much smaller in amplitude than the DH pairing, indicating that the traditional SC response is highly suppressed in the photodoped state. 
One fundamental difference is in the Coulomb interaction: it is attractive between a doublon and a holon, binding them into an exciton, while it is repulsive between two holons. 
Therefore, in a single rung, the DH state in the $\eta$-triplet sector is energetically favorable in the $\eta$-spin interaction term in Eq.~\eqref{eff} (see 
Supplemental Material~\cite{SM} for details).  

The emergence of long-range correlations along the chain is a nontrivial result that cannot be explained by local pair formation alone. 
Our DMRG calculations demonstrate that the DH pairing correlations can grow over long distances, revealing not only local exciton formation but also a long-range development of the pairing correlation.
The DH pairing state is stable in a wide regime of doublon density $n_d$ and Coulomb interaction $V$ as shown in Fig.~\ref{phase}(a)~\cite{ND}. 
In other regions of the $(n_d, V)$ phase diagram, various phases can dominate. 
Their detailed properties are presented in the Supplemental Material~\cite{SM}. 
Similar to the 1D chain~\cite{murakami2022, murakami2023}, the staggered on-site pairing, i.e., $\eta$-pairing, correlations become dominant in the small $V$ region, and the CDW correlations become dominant in the large $n_d$ and large $V$ region. 
At $n_d = 0$, we observe the spin-singlet phase, which is distinct from the SDW phase in the 1D chain, because a coupled spin chain with a spin gap~\cite{giamarchi2003} is stabilized.

\begin{figure}[t]
    \begin{center}       
        \includegraphics[width=\columnwidth]{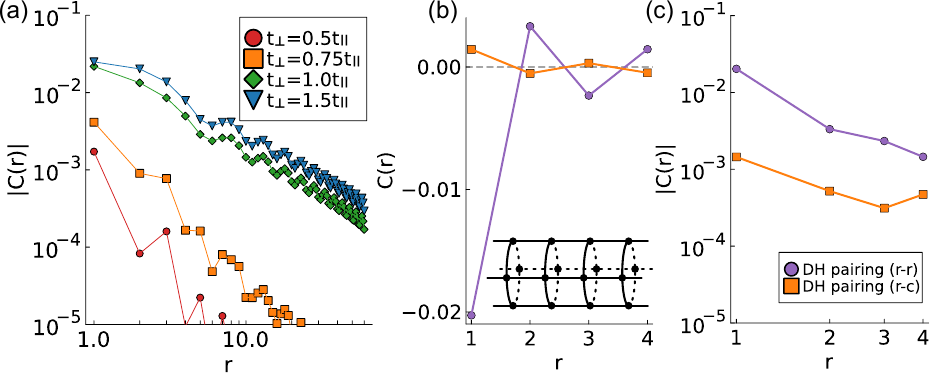}   
        \caption{(a) Rung-rung DH pairing correlation function for various $t_\perp / t_\parallel$ values at 
        $L = 120$,
        $n_d=0.1$ and $V = V_{\perp} = V_{\parallel} = 0.4t_{\parallel}$, where $J_\perp/J_\parallel = (t_{\perp}/t_{\parallel})^2$ with $J_{\parallel}=0.4t_{\parallel}$. 
        (b) Linear scale and (c) log-log plots of the DH pairing correlation functions in the four-leg cylinder, where $n_d=0.125$, $J = 0.4t_{\rm h}$, and $V = 0.8t_{\rm h}$. 
        }
        \label{phase2}
    \end{center}
\end{figure}

To examine the evolution of pairing tendency from 1D to ladder geometry, Fig.~\ref{phase2}(a) shows the DH pairing correlations for various $t_\perp/t_\parallel$ values. 
The DH pairing correlations increase as $t_\perp/t_\parallel$ increases. 
While the SDW correlation is dominant in the small-$t_\perp$ regime~\cite{SM}, the DH pairing correlation becomes dominant at  $t_\perp/t_\parallel \gtrsim 1$.  
This $t_\perp$ dependence suggests that DH pairing does not arise from one-dimensionality.
To extrapolate our observation towards 2D systems, we compute the DH pairing correlation on a cylinder square lattice with $8\times 4$ sites and periodic boundary conditions in the $y$ direction. 
Figures~\ref{phase2}(b) and \ref{phase2}(c) present the DH pairing correlations calculated by the DMRG method. 
Results on a cylinder are consistent with those on a ladder: we find a slow power-law decay of the pairing correlation, accompanied by the sign alternations. 
This suggests that DH pairing should not be limited to the ladder system and should persist in 2D.

\textit{Effective model}.
Having observed a stable staggered nature of the DH pairing, we derive here a minimal model that can explain the phase factor of $\braket{\hat{\Delta}^{\mathrm{r}\dagger}_{j+r}\hat{\Delta}^\mathrm{r}_{j}}$ in the ladder setup. 
The $t_\perp/t_\parallel$ dependence shown in Fig.~\ref{phase2}(a) suggests that a background of rung spin singlets, stabilized by the strong interchain coupling via $t_\perp$, is favorable for the emergence of the DH pairing phase on top of it. 
The preferable formation of rung spin singlets as $t_{\perp}/t_{\parallel}$ increases is confirmed in the Supplemental Material~\cite{SM}. 
Since the formation of DH $\eta$ triplets on the spin-singlet background is important for DH pairing characterized by the $\hat{\Delta}^{r}_j$ operator [see Fig.~\ref{phase}(d)], we discuss these behaviors in an effective model with local state space restricted to rung spin singlet and rung DH $\eta$ triplet. 

\begin{figure}[t]
    \begin{center}       
        \includegraphics[width=\columnwidth]{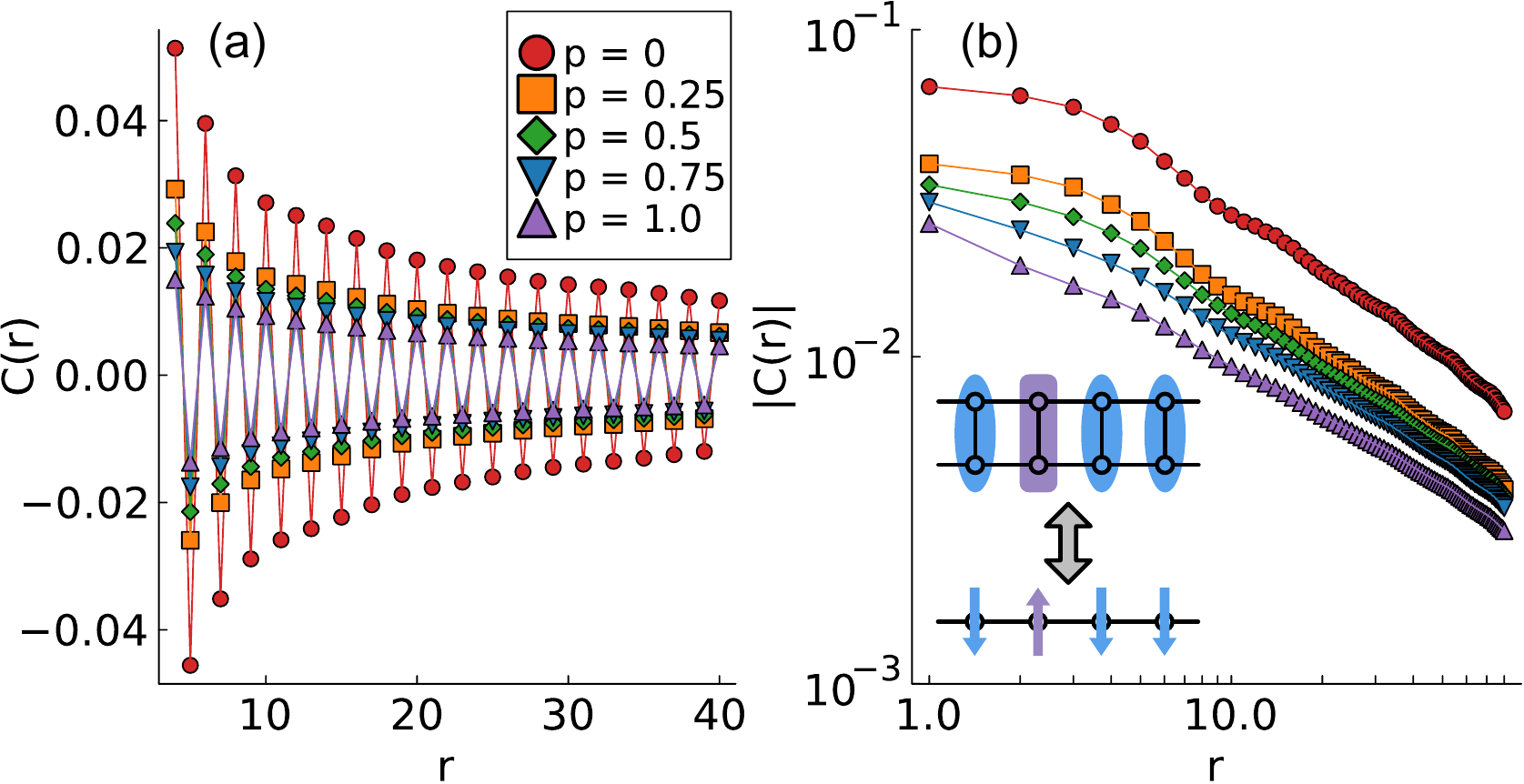}        
        \caption{(a) Linear scale plot and (b) log-log plot of the rung-rung DH pairing correlation function with varying $t_\parallel$, $J_\parallel$, and $V_\parallel$.
        Here, $t_\perp$ is used as the unit of energy.
        We set $n_d=0.05$, $J_\perp=0.4t_\perp$, and $V_\perp=1.8t_\perp$. 
        The interrung parameters are varied as $t_\parallel = (0.1+0.9p)t_\perp$, $J_\parallel = 0.4pt_\perp$, and $V_\parallel = 1.8pt_\perp$, where $0 \le p \le 1$. 
        Inset: Correspondence between the photodoped ladder and the 1D antiferromagnetic Heisenberg model. 
        The blue and purple symbols represent the rung spin singlet and rung DH $\eta$ triplet, respectively, as shown in Fig.~\ref{phase}(b).
        }
        \label{minimal}
    \end{center}
\end{figure}

We derive a minimal model from the local rung approximation~\cite{endres1996}, $\hat{\mathcal{H}}_{\rm LRA} = \hat{\mathcal{H}}_{\perp}+\hat{\mathcal{T}}_{\parallel}$, considering the strong rung coupling in $\hat{\mathcal{H}}_{\perp}$ [rung-direction component in Eq.~\eqref{eff}] with the interrung hopping $\hat{\mathcal{T}}_{\parallel}$ as the perturbation. 
The second-order perturbation theory gives the effective Hamiltonian for the two-level system  
\begin{equation}
\hat{\mathcal{H}}_{\rm min} = K
  \sum_{j} \left( \hat{\mathbf{\Delta}}_j \cdot \hat{\mathbf{\Delta}}_{j+1} - \frac{1}{4} \right), 
  \label{heisenberg}
\end{equation}
where $K = 2t_\parallel^2/(V_\perp + J_\perp + 2t_\perp ) + 2t_\parallel^2/(V_\perp + J_\perp - 2t_\perp )$ (see Supplemental Material~\cite{SM} for details). 
The rung DH $\eta$-triplet and rung spin-singlet states define the effective up and down pseudospin states, respectively [see the inset of Fig.~\ref{minimal}(b)].  
$\hat{\mathbf{\Delta}}_j$ is a vector composed of the pseudospin operators $\hat{\Delta}^+_j$, $\hat{\Delta}^-_j$, and $\hat{\Delta}^z_j$. 
As before, $\hat{\Delta}^+_j= \hat{\Delta}^{r\dagger}_j$ maps the rung spin-singlet state to the DH $\eta$-triplet state.  
The fixed number of doublons in a photodoped MI corresponds to the fixed pseudomagnetization in Eq.~(\ref{heisenberg}).  
Since this effective model is derived under the condition $K > 0$~\cite{SM}, Eq.~(\ref{heisenberg}) is equivalent to the 1D antiferromagnetic Heisenberg model at a certain magnetization. 
The lowest-energy state of Eq.~(\ref{heisenberg}) coincides with the ground state of the Heisenberg model under a magnetic field.
Due to the equivalence with the antiferromagnetic Heisenberg model, $\braket{\hat{\Delta}^{+}_{j+r}\hat{\Delta}^{-}_j}$ in Eq.~(\ref{heisenberg}) must have the phase factor $(-1)^r$, explaining the sign alternation of the DH pairing correlation $\braket{\hat{\Delta}^{\mathrm{r}\dagger}_{j+r}\hat{\Delta}^\mathrm{r}_j}$ in the photodoped ladder. 

Going beyond the above analytical perturbative arguments, we study the stability of the sign alternation in $\braket{\hat{\Delta}^{\mathrm{r}\dagger}_i\hat{\Delta}^\mathrm{r}_j}$ as a function of the $t_{\parallel}, J_{\parallel}, V_{\parallel}$ parameters of the full model $\hat{\mathcal{H}}$ numerically.
Figure~\ref{minimal} shows the DH pairing correlations as $t_\parallel$, $J_\parallel$, and $V_\parallel$ increase linearly, departing from $\hat{\mathcal{H}}_{\rm LRA}$.
As expected from the effective two-level model in Eq.~(\ref{heisenberg}), the DH pairing correlation is well developed when different rungs are only weakly connected ($p=0$). 
Although the absolute value of the correlation gradually decreases at larger $p$ as the parameters approach the original $\hat{\mathcal{H}}$, the sign alternation in the correlation function remains, confirming that $\hat{\mathcal{H}}_{\mathrm{min}}$ captures the essential features of the pairing correlations observed in the DH pairing phase. 

\textit{Conclusion}. Our DMRG calculations found the DH pairing phase, in which the correlation between DH pairs exhibits a sign alternation with $d$-wave-like symmetry. 
In the two-leg ladder, the DH pairing phase is found between the spin-singlet phase at $n_d\sim 0$ and the CDW/$\eta$-pairing phase in the large $n_d$ regime, suggesting that the interplay of charge, spin, and $\eta$-spin degrees of freedom in photodoped MIs can give rise to exotic quantum states analogous to unconventional SC states in chemically doped MIs. 
In contrast to the chemically doped one, the $\eta$-spin interaction, which is activated by photodoping, contributes to unconventional pairing.
Since the multileg cylinder of the 2D square lattice also exhibits the corresponding signature, the emergence of the DH pairing state is not specific to the two-leg ladder. 
Our results indicate the formation of DH pairs (Mott excitons) in a correlation-driven insulator, distinct from conventional electron-hole pairs in semiconductors. 
We further find a long-range development of the DH correlation, suggesting exciton condensation. 

The DH pairing state can appear in the small $n_d$ region, thus strong photoexcitation to achieve high $n_d$ is not required, and heating effects should be less destructive.
The Coulomb interaction $V$ cooperatively contributes to the emergence of the DH pairing state; such an interaction is natively present in many correlated materials.
For instance, ladder-type cuprates~\cite{ishida1994, azuma1994, eccleston1998, fukaya2015} may serve as hosts for the DH pairing state. 
Having charted the DH pairing in a 2D setting, various other materials, such as cuprates with the 2D square structure, hold the potential to exhibit DH pairing.
While the implementation of a nearest-neighbor Coulomb interaction is a challenging ingredient for the realization with cold atoms, the first step towards this has been achieved for the Bose-Hubbard model~\cite{weckesser2024}.

Although photodoping is inherently a nonequilibrium phenomenon, we have approximately mapped a photodoped state to the lowest-energy state in the pseudoequilibrium condition. 
To further validate the emergence of the DH pairing phase and its lifetime, it would be necessary to demonstrate the time-dependent behavior of the DH pairing correlations in the optically driven Hubbard ladder using time-evolution methods~\cite{vidal2004, vidal2007, haegeman2011, haegeman2016, paeckel2019}. 
Having obtained a promising behavior on a small 2D cylinder, further studies of the $d$-wave nature of pairing should be performed on larger 2D lattices.
These investigations remain as future challenges.

\textit{Acknowledgments}.
We thank S.~Ejima, Y.~Murakami, S.~Nishimoto, K.~Pradhan, and T.~Sato for fruitful discussions.
This work was supported by Grants-in-Aid for Scientific Research from JSPS, KAKENHI Grants No.~JP20H01849 (T.K.), No.~JP24K06939 (T.K.), No.~JP24H00191 (T.K.), No.~JP22K04907 (K.K.), and No.~JP24K01333. 
R.U. was supported by the Program for Leading Graduate Schools: ``Interactive Materials Science Cadet Program'' and JST SPRING Grant No.~JPMJSP2138. 
M.S. acknowledges the support from the U.K. EPSRC award under Agreement No.~EP/Y005090/1. 
Z.L. acknowledges the support by the QuantERA II JTC 2021 Grant T-NiSQ by MVZI, the P1-0044 program of the Slovenian Research Agency, and ERC StG 2022 project DrumS, Grant Agreement No.~101077265. 
D.G. is supported by the Slovenian Research and Innovation Agency (ARIS) under Programs No.~P1-0044, No.~J1-2455, and No.~MN-0016-106. 
Our calculations were performed using the ITensor library~\cite{ITensor,ITensor-r0.3}.

\textit{Data availability}. 
The data that support the findings of this article are openly available~\cite{data}.

\bibliography{Reference}

\begin{thebibliography}{112}
\bibitem [{zig()}]{zigzag}%
  \BibitemOpen
  \href@noop {} {}\bibinfo {note} {To perform the DMRG calculations, we map the ladder and cylinder geometries onto a 1D chain using a zigzag path.}\BibitemShut {Stop}%
\bibitem [{CF()}]{CF}%
  \BibitemOpen
  \href@noop {} {}\bibinfo {note} {Because the correlations between these single-site operators are symmetric on both chains, we consider the $\alpha=0$ chain index in the operators.}\BibitemShut {Stop}%
\bibitem [{ND()}]{ND}%
  \BibitemOpen
  \href@noop {} {}\bibinfo {note} {At $n_d=0.25$, the DH pairing correlations exhibited a slightly exponential trend rather than a power-law decay; however, DH pairing was the most developed among the correlation functions we examined.}\BibitemShut {Stop}%
\end{thebibliography}

\end{document}


\title{Supplemental Material: \\
Doublon-holon pairing state in photodoped Mott insulators}
\author{
Ryota Ueda$^1$, 
Madhumita Sarkar$^{2,3}$,
Zala Lenar\v{c}i\v{c}$^2$,
Denis Gole\v{z}$^{2,4}$,
Kazuhiko Kuroki$^1$, and
Tatsuya Kaneko$^1$
}
\affiliation{
$^1$Department of Physics, The University of Osaka, Toyonaka, Osaka 560-0043, Japan \\
$^2$Jo\v{z}ef Stefan Institute, Jamova 39, SI-1000 Ljubljana, Slovenia \\
$^3$Department of Physics and Astronomy, University of Exeter, Stocker Road, Exeter EX4 4QL, United Kingdom\\
$^4$Faculty of Mathematics and Physics, University of Ljubljana, Jadranska 19, 1000 Ljubljana, Slovenia
}
\date{\today}
\maketitle

\renewcommand{\thefigure}{S\arabic{figure}}

\section{Hamiltonian for Photodoped Ladder-type Mott Insulators}

The Hamiltonian of the extended Hubbard ladder is given by
\begin{align} 
  \label{eq: ham}
  \hat{\mathcal{H}}_{\mathrm{Hub}} =
  &- t_\parallel \sum_{j, \alpha, \sigma} 
  \left( \hat{c}^\dagger_{j, \alpha; \sigma} \hat{c}_{j+1, \alpha; \sigma} + \mathrm{H.c.} \right) 
  - t_\perp \sum_{j, \sigma} 
  \left( \hat{c}^\dagger_{j, 0; \sigma} \hat{c}_{j, 1; \sigma} + \mathrm{H.c.} \right)
  \nonumber \\
  &+ U \sum_{j, \alpha}
  \left( \hat{n}_{j, \alpha; \uparrow} - \frac{1}{2} \right)
  \left( \hat{n}_{j, \alpha; \downarrow} - \frac{1}{2} \right)
  + V_\parallel \sum_{j, \alpha} 
  \left( \hat{n}_{j, \alpha} - 1 \right)
  \left( \hat{n}_{j+1, \alpha} - 1 \right)
  + V_\perp \sum_{j} 
  \left( \hat{n}_{j, 0} - 1 \right)
  \left( \hat{n}_{j, 1} - 1 \right).
\end{align}
$\hat{c}^\dagger_{j, \alpha; \sigma}$ ($\hat{c}_{j, \alpha; \sigma}$) is the creation (annihilation) operator for a fermion with spin $\sigma = \uparrow$, $\downarrow$ at site $j$ on chain $\alpha \left( =0,1 \right)$. 
$\hat{n}_{j, \alpha; \sigma} = \hat{c}^\dagger_{j, \alpha; \sigma} \hat{c}_{j, \alpha; \sigma}$ is the number operator, and $\hat{n}_{j, \alpha}=\hat{n}_{j, \alpha; \uparrow} + \hat{n}_{j, \alpha; \downarrow}$.
$t_\parallel$ and $t_\perp$ are the hopping integrals along the chain and rung directions, respectively.
$U > 0$ is the on-site Coulomb interaction.
$V_\parallel > 0$ and $V_\perp > 0$ are the nearest-neighbor Coulomb interactions along the chain and rung directions, respectively. 

\begin{figure}[t]
    \centering
        \includegraphics[width=0.55\columnwidth]{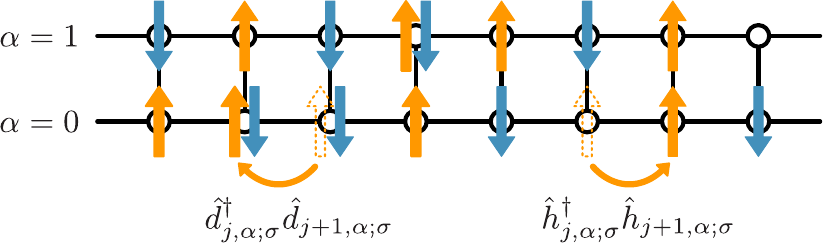}
        \caption{Schematic diagram of doublon-number-conserving hoppings. 
        The operators $\hat{d}^\dagger_{j,\alpha,\sigma} \hat{d}_{j+1,\alpha,\sigma}$ and $\hat{h}^\dagger_{j,\alpha,\sigma} \hat{h}_{j+1,\alpha,\sigma}$ transfer a doublon and a holon to an adjacent site, respectively, without changing the number of doublons. 
        }
        \label{hopping}
\end{figure}

\begin{figure}[t]
    \centering
        \includegraphics[width=\columnwidth]{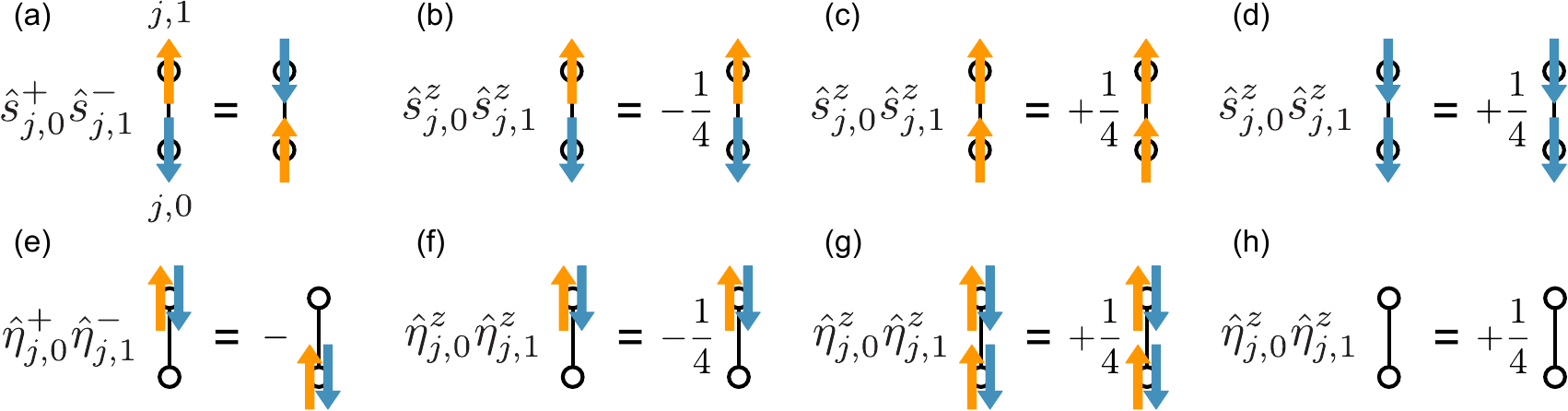}
        \caption{Schematic diagrams of the spin ($\hat{s}$) and $\eta$-spin ($\hat{\eta}$) interactions between sites $\bm{j}=(j,0)$ and $(j,1)$. 
        (a)-(d) Spin exchange term $\hat{s}^+_{j,0}\hat{s}^-_{j,1}$ and $zz$ term $\hat{s}^z_{j,0}\hat{s}^z_{j,1}$ that act on singly occupied (up or down) sites.
        (e)-(h) $\eta$-spin exchange term $\hat{\eta}^+_{j,0}\hat{\eta}^-_{j,1}$ and $zz$ term $\hat{\eta}^z_{j,0}\hat{\eta}^z_{j,1}$ that act on doublon and holon sites.
        }
        \label{interaction}
\end{figure}

We consider the effective $t$-$J$-$V$ model for photodoped Mott insulators (MIs) in the strong-coupling ($U \gg t_{\parallel}, t_{\perp}$) limit~\cite{li2020,murakami2022,murakami2023}.
The effective model for the ladder system is constructed as~\cite{sarkar2024}
\begin{align}
  \hat{\mathcal{H}} &= 
  \hat{\mathcal{H}}^{(0)}_{t}
  +\hat{\mathcal{H}}^{(s)}_{J}
  +\hat{\mathcal{H}}^{(\eta)}_{J}
  +\hat{\mathcal{H}}_{V} \nonumber \\
  &= \hat{\mathcal{H}}^{(0)}_{t_\parallel} + \hat{\mathcal{H}}^{(0)}_{t_\perp}
  + \hat{\mathcal{H}}^{(s)}_{J_\parallel} + \hat{\mathcal{H}}^{(s)}_{J_\perp}
  + \hat{\mathcal{H}}^{(\eta)}_{J_\parallel} + \hat{\mathcal{H}}^{(\eta)}_{J_\perp} 
    + \hat{\mathcal{H}}_{V_\parallel} + \hat{\mathcal{H}}_{V_\perp}.
  \label{eff}
\end{align}
$\hat{\mathcal{H}}^{(0)}_{t_\parallel}$ and $\hat{\mathcal{H}}^{(0)}_{t_\perp}$ represent the doublon-number-conserving hopping along the chain and rung directions, respectively:
\begin{align}
& \hat{\mathcal{H}}^{(0)}_{t_\parallel} = - t_\parallel \sum_{j, \alpha, \sigma} 
  \left( \left( 1 - \hat{n}_{j, \alpha; \bar{\sigma}} \right) \hat{c}^\dagger_{j, \alpha; \sigma} \hat{c}_{j+1, \alpha; \sigma}  \left( 1 - \hat{n}_{j+1, \alpha; \bar{\sigma}} \right) + \hat{n}_{j, \alpha; \bar{\sigma}} \hat{c}^\dagger_{j, \alpha; \sigma} \hat{c}_{j+1, \alpha; \sigma}  \hat{n}_{j+1, \alpha; \bar{\sigma}} + \mathrm{H.c.} \right), 
  \\
& \hat{\mathcal{H}}^{(0)}_{t_\perp} = 
  - t_\perp \sum_{j, \sigma} 
  \left(  \left( 1 - \hat{n}_{j, 0; \bar{\sigma}} \right)  \hat{c}^\dagger_{j, 0; \sigma} \hat{c}_{j, 1; \sigma} \left( 1 - \hat{n}_{j, 1; \bar{\sigma}} \right) + \hat{n}_{j, 0; \bar{\sigma}} \hat{c}^\dagger_{j, 0; \sigma} \hat{c}_{j, 1; \sigma} \hat{n}_{j, 1; \bar{\sigma}} + \mathrm{H.c.} \right). 
\end{align}
In the main text, these hopping terms are described using the operators $\hat{h}^\dagger_{j,\alpha;\sigma}=\hat{c}_{j,\alpha;\sigma}(1-\hat{n}_{j,\alpha;\bar{\sigma}})$ for a holon and $\hat{d}^\dagger_{j,\alpha;\sigma}=\hat{c}^\dagger_{j,\alpha;\sigma}\hat{n}_{j,\alpha;\bar{\sigma}}$ for a doublon, where $\bar{\sigma}=\downarrow$ $(\uparrow)$ for $\sigma=\uparrow$ $(\downarrow)$.
Figure~\ref{hopping} schematically shows the doublon-number-conserving hopping $\hat{\mathcal{H}}^{(0)}_{t_\parallel}$ described by 
$\hat{d}^\dagger_{j,\alpha,\sigma} \hat{d}_{j+1,\alpha,\sigma}$ and $\hat{h}^\dagger_{j,\alpha,\sigma} \hat{h}_{j+1,\alpha,\sigma}$.
$\hat{\mathcal{H}}^{(s)}_{J_\parallel}$ and $\hat{\mathcal{H}}^{(s)}_{J_\perp}$ represent the Heisenberg-type spin interactions along the chain and rung directions, respectively:
\begin{align}
&  \hat{\mathcal{H}}^{(s)}_{J_\parallel} = 
  J_{\parallel} \sum_{j, \alpha} \left( \hat{\bm{s}}_{j, \alpha} \cdot \hat{\bm{s}}_{j+1, \alpha} - \frac{1}{4} 
  \delta_{1,\hat{n}_{j, \alpha}}\delta_{1,\hat{n}_{j+1, \alpha}}
  \right), 
  \\
&  \hat{\mathcal{H}}^{(s)}_{J_\perp} = 
  J_{\perp} \sum_{j} \left( \hat{\bm{s}}_{j, 0} \cdot \hat{\bm{s}}_{j, 1} - \frac{1}{4} 
  \delta_{1,\hat{n}_{j, 0}}\delta_{1,\hat{n}_{j, 1}} 
  \right).  
\end{align}
$\hat{\bm{s}}_{j,\alpha} = \sum_{\sigma, \sigma'} \hat{c}^\dagger_{j,\alpha;\sigma} \bm{\sigma}_{\sigma\sigma'} \hat{c}_{j,\alpha;\sigma'} / 2$ is the spin operator, where $\bm{\sigma}$ is the vector of Pauli matrices.
$\delta_{1,\hat{n}_{j, \alpha}}\delta_{1,\hat{n}_{j', \alpha'}} = 1$ 
only when both adjacent sites [$(j,\alpha)$ and $(j',\alpha')$] are the singly occupied (up or down) sites.
These spin interactions act between singly occupied sites as illustrated in Figs.~\ref{interaction}(a)-\ref{interaction}(d).
The coupling constants are given by $J_{\parallel} =4t^2_{\parallel}/U$ and $J_{\perp} =4t^2_{\perp}/U$ (at $V_{\parallel}=V_{\perp}=0$). 
$\hat{\mathcal{H}}^{(\eta)}_{J_\parallel}$ and $\hat{\mathcal{H}}^{(\eta)}_{J_\perp}$ describe the $\eta$-spin interactions along the chain and rung directions, respectively:
\begin{align}
&  \hat{\mathcal{H}}^{(\eta)}_{J_\parallel} = 
  - J_{\parallel} \sum_{j, \alpha} \left( \hat{\bm{\eta}}_{j, \alpha} \cdot \hat{\bm{\eta}}_{j+1, \alpha} - \frac{1}{4} \left( 1 - \delta_{1,\hat{n}_{j, \alpha}} \right) \left( 1 - \delta_{1,\hat{n}_{j+1, \alpha}} \right) \right), 
  \\
&  \hat{\mathcal{H}}^{(\eta)}_{J_\perp} = 
  - J_{\perp} \sum_{j} \left( \hat{\bm{\eta}}_{j, 0} \cdot \hat{\bm{\eta}}_{j, 1} - \frac{1}{4} \left( 1 - \delta_{1,\hat{n}_{j, 0}} \right) \left( 1 - \delta_{1,\hat{n}_{j, 1}} \right) \right).
\end{align}
The $\eta$-spin operator $\hat{\bm{\eta}}_{j,\alpha}$ is given by $\hat{\eta}^+_{j,\alpha} = (-1)^{j+\alpha} \hat{c}^\dagger_{j,\alpha; \downarrow} \hat{c}^\dagger_{j,\alpha; \uparrow}$, $\hat{\eta}^-_{j,\alpha} = (-1)^{j+\alpha} \hat{c}_{j,\alpha; \uparrow} \hat{c}_{j,\alpha; \downarrow}$, and $\hat{\eta}^z_{j,\alpha} = \left( \hat{n}_{j,\alpha} - 1\right) / 2$~\cite{essler2005}. 
$\delta_{1,\hat{n}_{j, \alpha}}=1$ only when the site $(j,\alpha)$ is the singly occupied (up or down) site, 
whereas $1-\delta_{1,\hat{n}_{j, \alpha}}=1$ when that is empty or doubly occupied (holon or doublon). 
In the main text, we introduce $\bar{\delta}_{1,\hat{n}_{j, \alpha}} = 1-\delta_{1,\hat{n}_{j, \alpha}}$ to simplify the notation. 
These $\eta$-spin interactions for empty/doubly occupied sites are illustrated in Figs.~\ref{interaction}(e)-\ref{interaction}(h). 
$\hat{\eta}^{+}_{j, 0} \hat{\eta}^{-}_{j, 1}$ swaps the positions of a doublon and a holon, while $\hat{\eta}^{z}_{j, 0} \hat{\eta}^{z}_{j, 1}$
measures the energy according to the occupancy status of two sites. 
Similarly to the spin interaction shown in Figs.~\ref{interaction}(a)-\ref{interaction}(d), the $\eta$-spin interaction acts as an exchange interaction between doublons and holons. 
$\hat{\mathcal{H}}_{V_\parallel}$ and $\hat{\mathcal{H}}_{V_\perp}$ describe the nearest-neighbor Coulomb interactions along the chain and rung directions, respectively:
\begin{align}
&  \hat{\mathcal{H}}_{V_\parallel} =
  V_\parallel \sum_{j, \alpha} \left( \hat{n}_{j, \alpha} - 1 \right) \left( \hat{n}_{j+1, \alpha} - 1 \right)
  = 4V_\parallel \sum_{j, \alpha} \hat{\eta}^z_{j, \alpha} \hat{\eta}^z_{j+1, \alpha} , 
  \\ 
&  \hat{\mathcal{H}}_{V_\perp} =
  V_\perp \sum_{j} \left( \hat{n}_{j, 0} - 1 \right) \left( \hat{n}_{j, 1} - 1 \right) 
  = 4V_\perp \sum_{j} \hat{\eta}^z_{j, 0} \hat{\eta}^z_{j, 1} . 
\end{align}
As shown above, the nearest-neighbor Coulomb interaction term can be described using the operator $\hat{\eta}^z_{j,\alpha}$. 

\section{Pair Operators}

\begin{figure}[b]
    \centering
        \includegraphics[width=\columnwidth]{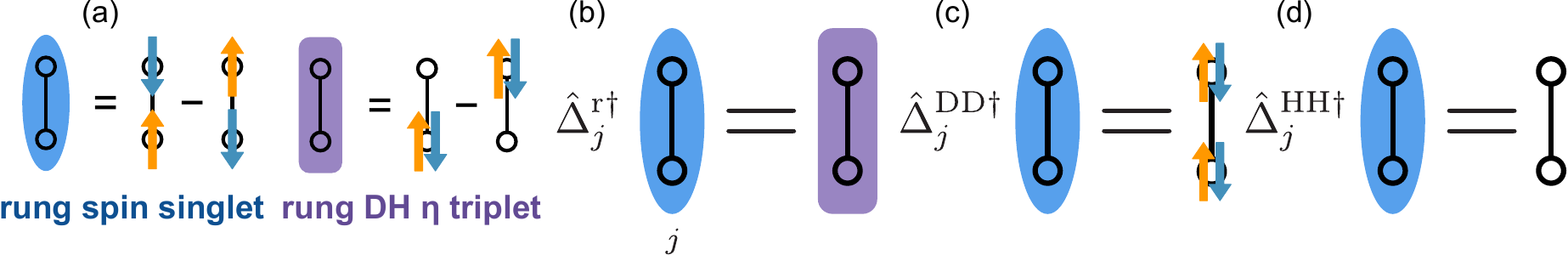}
        \caption{(a) Schematic diagrams of the rung spin singlet and the rung DH $\eta$ triplet.
        Schematic diagrams showing the definitions of the rung (b) DH pair, (c) DD pair, and (d) HH pair operators.  
        These operators map the rung spin-singlet state to the rung (b) DH, (c) DD, and (d) HH $\eta$-triplet states, respectively.
        }
        \label{operator}
\end{figure}

We summarize the definitions of the pair operators composed of doublons and holons, which act as carriers in the photodoped state.
The rung DH pair operator is defined as
\begin{equation}
    \hat{\Delta}^{\mathrm{r}\dagger}_{j} = \frac{1}{2} \sum_{\alpha} \sum_{\sigma} (-1)^{\alpha} \hat{n}_{j, \alpha; \bar{\sigma}} \hat{c}^\dagger_{j, \alpha;\sigma} \hat{c}_{j, \bar{\alpha};\sigma} \left( 1 - \hat{n}_{j, \bar{\alpha}; \bar{\sigma}} \right).
\end{equation} 
The operator $\hat{\Delta}^\mathrm{r}_{j}$ satisfies the relation
\begin{equation} \label{r-r_DH}
    \hat{\Delta}^{\mathrm{r}\dag}_{j} \ket{s}_j = \ket{\eta=1, \eta^z=0}_j.  
\end{equation}
Here, $\ket{s}_j$ denotes the rung spin-singlet state  
\begin{align}
    &\ket{s}_j = \frac{1}{\sqrt{2}}\left(\hat{c}^\dagger_{j,0;\uparrow}
     \hat{c}^\dagger_{j,1;\downarrow} - \hat{c}^\dagger_{j,0;\downarrow} \hat{c}^\dagger_{j,1;\uparrow}\right) \ket{0}_j,
\end{align}
and $\ket{\eta=1, \eta^z=0}_j$ represents the rung $\eta$-triplet state composed of a doublon (D) and a holon (H) 
\begin{align}
   &\ket{\eta=1, \eta^z=0}_j = \frac{1}{\sqrt{2}}\left(\hat{c}^\dagger_{j,0;\uparrow} \hat{c}^\dagger_{j,0;\downarrow} - \hat{c}^\dagger_{j,1;\uparrow} \hat{c}^\dagger_{j,1;\downarrow}\right) \ket{0}_j,
\end{align}
where $\ket{0}_j$ denotes the vacuum state at the $j$-th rung.
The rung spin-singlet state and the rung DH $\eta$-triplet state are schematically shown in Fig.~\ref{operator}(a).
$\eta$ and $\eta^z$ represent the quantum numbers for $\eta$ spins on a single rung characterized by 
$\hat{\bm{\eta}}^2_j\ket{\eta, \eta^z}_j=\eta(\eta+1)\ket{\eta, \eta^z}_j$ and $\hat{\eta}^z_j\ket{\eta, \eta^z}_j=\eta^z\ket{\eta, \eta^z}_j$, where $\hat{\bm{\eta}}_j = \sum_{\alpha}\hat{\bm{\eta}}_{j,\alpha}$.
As shown in Fig.~\ref{operator}(b), $\hat{\Delta}^{\mathrm{r}\dagger}_{j}$ maps the rung spin singlet to the rung DH $\eta$ triplet.
The DH pair formed along the chain direction is described by 
\begin{equation}
    \hat{\Delta}^{\mathrm{c}\dagger}_{j} = \frac{1}{2} \sum_{\beta=0,1} \sum_{\sigma} (-1)^{\beta} \hat{n}_{j+\beta, 0; \bar{\sigma}} \hat{c}^\dagger_{j+\beta, 0;\sigma} \hat{c}_{j+\bar{\beta}, 0;\sigma} \left( 1 - \hat{n}_{j+\bar{\beta}, 0; \bar{\sigma}} \right).
    \label{chaindh}
\end{equation}
Using these operators, the rung-rung and rung-chain DH pairing correlation functions are defined as $\braket{\hat{\Delta}^{\mathrm{r}\dagger}_{j_0+r}\hat{\Delta}^\mathrm{r}_{j_0}}$ and $\braket{\hat{\Delta}^{\mathrm{c}\dagger}_{j_0+r}\hat{\Delta}^\mathrm{r}_{j_0}}$, respectively.
Note that the operators $\hat{h}^\dagger_{j,\alpha;\sigma}=\hat{c}_{j,\alpha;\sigma}(1-\hat{n}_{j,\alpha;\bar{\sigma}})$ and $\hat{d}^\dagger_{j,\alpha;\sigma}=\hat{c}^\dagger_{j,\alpha;\sigma}\hat{n}_{j,\alpha;\bar{\sigma}}$ are used to describe $\hat{\Delta}^{\mathrm{r}\dag}_{j}$ and $\hat{\Delta}^{\mathrm{c}\dag}_{j}$ in the main text.

\begin{figure}[t]
    \centering
        \includegraphics[width=0.5\columnwidth]{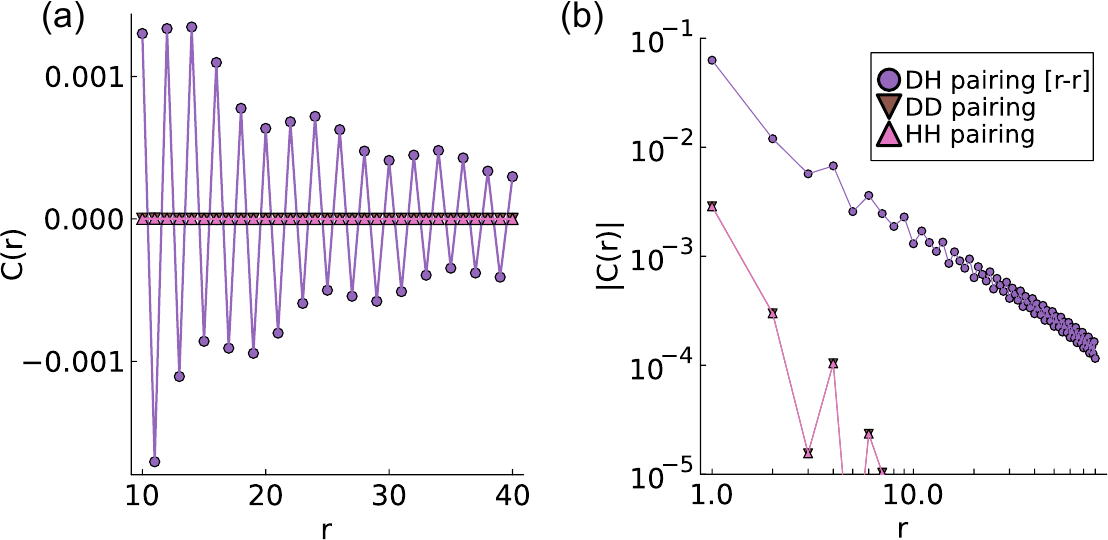}
        \caption{(a) Linear scale and (b) log-log plots of the pairing correlation functions in the DH pairing phase ($L = 160$, $n_d = 0.2$, $t_\perp = t_\parallel$, $J_{\parallel} = J_{\perp} = 0.4t_\parallel$, and $V = V_{\perp} = V_{\parallel} = 0.2t_\parallel$).
The purple, brown, and pink points represent the rung-rung DH, DD, and HH pairing correlations, respectively.
The DD and HH pairing correlations exhibit degeneracy.
        }
        \label{ddhh}
\end{figure}

Meanwhile, the doublon-doublon (DD) pair operator is defined as 
\begin{align}
\hat{\Delta}_j^{\mathrm{DD}\dagger}
= \frac{1}{\sqrt{2}} \sum_{\alpha} 
\hat{n}_{j, \alpha; \uparrow}
\hat{c}^\dagger_{j, \alpha; \downarrow} 
\hat{c}^\dagger_{j, \bar{\alpha}; \uparrow} 
\hat{n}_{j, \bar{\alpha}; \downarrow} 
\end{align}
which satisfies the relation
\begin{align}
    \hat{\Delta}_j^{\mathrm{DD}\dagger} \ket{s}_j = \ket{\eta=1, \eta^z=1}_j
\end{align}
for the $\eta$-triplet state composed of two doublons
\begin{equation}
    \ket{\eta=1, \eta^z=1}_j = \hat{c}_{j,0; \uparrow}^{\dagger} \hat{c}_{j,0; \downarrow}^{\dagger} \hat{c}_{j,1; \uparrow}^{\dagger} \hat{c}_{j,1; \downarrow}^{\dagger} \ket{0}_j. 
\end{equation}
The holon-holon (HH) pair operator is defined as
\begin{align}
\hat{\Delta}_j^{\mathrm{HH}\dagger}
= \frac{1}{\sqrt{2}} \sum_{\alpha} 
\left(1 - \hat{n}_{j, \alpha; \uparrow}\right)
\hat{c}_{j, \alpha; \downarrow}
\hat{c}_{j, \bar{\alpha}; \uparrow}
\left(1 - \hat{n}_{j, \bar{\alpha}; \downarrow}\right) 
\end{align}
which satisfies the relation
\begin{align}
    \hat{\Delta}_j^{\mathrm{HH}\dagger} \ket{s}_j = \ket{\eta=1, \eta^z=-1}_j
\end{align}
for the $\eta$-triplet state composed of two holons
\begin{equation}
    \ket{\eta=1, \eta^z=-1}_j = \ket{0}_j.
\end{equation}
As shown in Figs.~\ref{operator}(c) and \ref{operator}(d), $\hat{\Delta}_j^{\mathrm{DD}\dagger}$ and $\hat{\Delta}_j^{\mathrm{HH}\dagger}$ map the rung spin-singlet state to the rung DD and HH $\eta$-triplet states, respectively. 
Using these operators, the DD and HH pairing correlation functions are defined as $\braket{\hat{\Delta}^{\mathrm{DD}\dagger}_{j_0+r}\hat{\Delta}^{\mathrm{DD}}_{j_0}}$ and $\braket{\hat{\Delta}^{\mathrm{HH}\dagger}_{j_0+r}\hat{\Delta}^{\mathrm{HH}}_{j_0}}$, respectively. 
Figure~\ref{ddhh} shows the DH, DD, and HH pairing correlation functions in the DH pairing phase.  
Both DD and HH pairing correlations are much weaker than the DH pairing correlation. 

To understand why the DD and HH pairing correlations are weaker, we consider a two-site model on a single rung given by 
\begin{align}
  \hat{\mathcal{H}}_{\mathrm{rung}} 
  =   \hat{\mathcal{H}}^{(0)}_{\mathrm{rung}, t_\perp}
  +\hat{\mathcal{H}}^{(s)}_{\mathrm{rung}, J_\perp}
  +\hat{\mathcal{H}}^{(\eta)}_{\mathrm{rung}, J_\perp}
  +\hat{\mathcal{H}}_{\mathrm{rung}, V_\perp}
  \label{eq:H_rung}
\end{align}
with 
\begin{align}
&  \hat{\mathcal{H}}^{(0)}_{\mathrm{rung}, t_\perp} = -t_\perp \sum_{\sigma} \left( (1 - \hat{n}_{0; \bar{\sigma}}) \hat{c}^\dagger_{0; \sigma} \hat{c}_{1; \sigma} (1 - \hat{n}_{1; \bar{\sigma}}) + \hat{n}_{0; \bar{\sigma}} \hat{c}^\dagger_{0; \sigma} \hat{c}_{1; \sigma} \hat{n}_{1; \bar{\sigma}} + \mathrm{H.c.} \right),
  \\
&  \hat{\mathcal{H}}^{(s)}_{\mathrm{rung}, J_\perp} = J_\perp \left(\hat{\bm{s}}_0 \cdot \hat{\bm{s}}_1 - \frac{1}{4} \delta_{1,\hat{n}_0} \delta_{1,\hat{n}_1}\right),
\\
&  \hat{\mathcal{H}}^{(\eta)}_{\mathrm{rung}, J_\perp} + \hat{\mathcal{H}}_{\mathrm{rung}, V_\perp}
  = - J_\perp \left(\hat{\bm{\eta}}_0 \cdot \hat{\bm{\eta}}_1 - \frac{1}{4} \left(1 - \delta_{1,\hat{n}_0}\right)\left(1 - \delta_{1,\hat{n}_1}\right)\right)
  + 4V_\perp \hat{\eta}_0^z \hat{\eta}_1^z. 
\end{align}
When only doublons and holons are present, $\hat{\mathcal{H}}^{(0)}_{\mathrm{rung}, t_\perp}$ and $\hat{\mathcal{H}}^{(s)}_{\mathrm{rung}, J_\perp}$ do not contribute.
Hence, we consider the eigenstates of $\hat{\mathcal{H}}^{(\eta)}_{\mathrm{rung}, J_\perp} + \hat{\mathcal{H}}_{\mathrm{rung}, V_\perp}$. 
In the two-site model, the eigenstate and eigenenergy for the $\eta=0$ state are given by
\begin{align}
    &\ket{\eta=0, \eta^z=0} = \frac{1}{\sqrt{2}}\left(\hat{c}^\dagger_{0;\uparrow} \hat{c}^\dagger_{0;\downarrow} + \hat{c}^\dagger_{1;\uparrow} \hat{c}^\dagger_{1;\downarrow}\right) \ket{0}, 
    \quad \hat{\mathcal{H}}_{\mathrm{rung}} \ket{\eta=0, \eta^z=0} = (-V_\perp + J_\perp) \ket{\eta=0, \eta^z=0},
\end{align}
while those for the $\eta=1$ states are given by 
\begin{align}
    &\ket{\eta=1, \eta^z=1} = \hat{c}_{0; \uparrow}^{\dagger} \hat{c}_{0; \downarrow}^{\dagger} \hat{c}_{1; \uparrow}^{\dagger} \hat{c}_{1; \downarrow}^{\dagger} \ket{0}, 
    \hspace{51pt} \hat{\mathcal{H}}_{\mathrm{rung}} \ket{\eta=1, \eta^z=1} = V_\perp \ket{\eta=1, \eta^z=1}, 
    \\
    &\ket{\eta=1, \eta^z=0} = \frac{1}{\sqrt{2}}\left(\hat{c}^\dagger_{0;\uparrow} \hat{c}^\dagger_{0;\downarrow} - \hat{c}^\dagger_{1;\uparrow} \hat{c}^\dagger_{1;\downarrow}\right) \ket{0}, 
    \quad \hat{\mathcal{H}}_{\mathrm{rung}} \ket{\eta=1, \eta^z=0} = -V_\perp \ket{\eta=1, \eta^z=0}, 
    \\
    &\ket{\eta=1, \eta^z=-1} = \ket{0}, 
    \hspace{106pt}  \hat{\mathcal{H}}_{\mathrm{rung}} \ket{\eta=1, \eta^z=-1} = V_\perp \ket{\eta=1, \eta^z=-1}. 
\end{align}
$\ket{\eta=0, \eta^z=0}$ represents the $\eta$ singlet composed of a DH pair, while $\ket{\eta=1, \eta^z=1}$, $\ket{\eta=1, \eta^z=0}$, and $\ket{\eta=1, \eta^z=-1}$ correspond to the $\eta$ triplets composed of DD, DH, and HH pairs, respectively. 
Note that the presence of two doublons in a single rung means the presence of two excess holons in other places, and the on-site energy of the whole photodoped system is constant. 
When $V_{\perp}=0$, the energies of $\ket{\eta=1, \eta^z=1}$, $\ket{\eta=1, \eta^z=0}$, and $\ket{\eta=1, \eta^z=-1}$ are degenerate and lower than than the energy of $\ket{\eta=0, \eta^z=0}$ because the $\eta$-spin interaction is ferromagnetic.  
The repulsive interaction $V_{\perp}$ lifts the energy degeneracy of the $\eta$-triplets states, and the lowest-energy state becomes $\ket{\eta=1, \eta^z=0}$.
Therefore, the DH $\eta$-triplet state is energetically favored in a single rung.  

In the absence of doublons and holons on a rung, the spin-singlet state is the ground state of the two-site antiferromagnetic Heisenberg model. 
As the photoexcited carriers relax into a quasisteady lowest-energy state, rungs with singly occupied fermions are predominantly in the spin-singlet state $\ket{s}_j$, while rungs with a DH pair are predominantly in the $\eta$-triplet state $\ket{\eta=1, \eta^z=0}$. 
These energetic arguments motivate our choice of the pairing operator $\hat{\Delta}^{\mathrm{r}}_{j}$, which connects these low-energy configurations. 

\section{Supplemental Numerical Results} 

\subsection{Verification of Numerical Accuracy}

\begin{figure}[b]
  \begin{minipage}[t]{\columnwidth}
      \centering
      \includegraphics[width=\columnwidth]{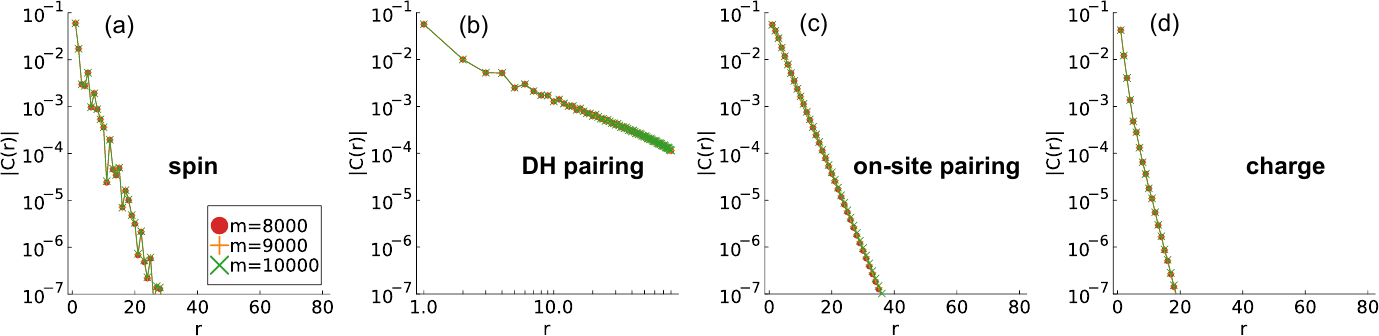}
      \caption{(a) Spin, (b) rung-rung DH pairing, (c) on-site pairing, and (d) charge correlation functions for $t_\perp = t_\parallel$, where $L=160$, $n_d = 0.2$, $J_{\perp} = J_{\parallel}= 0.4t_\parallel$, and $V = V_{\perp} = V_{\parallel} = 0.2t_\parallel$. 
      The data in (b) are presented in a log-log plot. 
      }
      \label{1}
  \end{minipage}
  
  \begin{minipage}[t]{\columnwidth}
      \centering
      \includegraphics[width=\columnwidth]{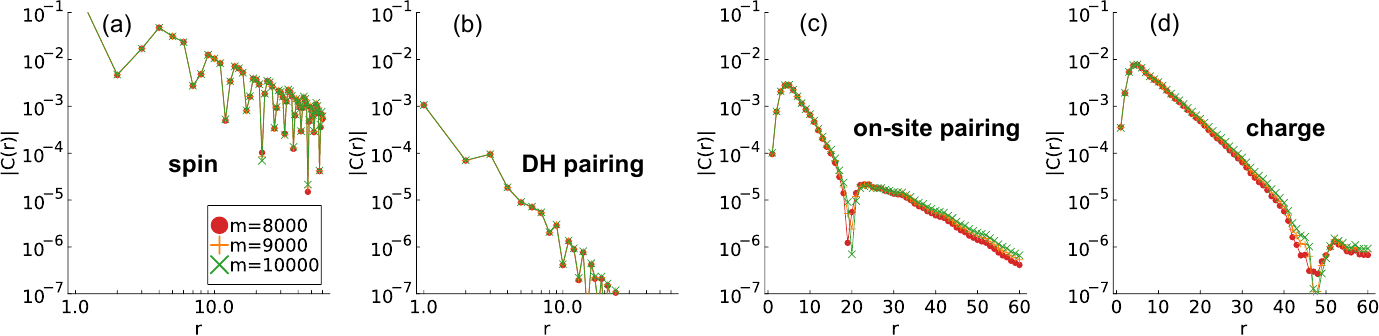}
      \caption{(a) Spin, (b) rung-rung DH pairing, (c) on-site pairing, and (d) charge correlation functions for $t_\perp = 0.25t_\parallel$, where $L=120$, $n_d = 0.1$, $J_\parallel = 0.4t_\parallel$, $J_\perp = (t_{\perp}/t_{\parallel})^2J_\parallel$, and $V = V_{\perp} = V_{\parallel} = 0.2t_\parallel$.
      The data in (a) and (b) are presented in log-log plots.   
      }
      \label{0.25}
  \end{minipage}
\end{figure}

Figures \ref{1} and \ref{0.25} show the correlation functions for $t_\perp = t_\parallel$ and $t_\perp = 0.25t_\parallel$, respectively, to confirm the numerical accuracy of the results presented in Figs. 2 and 3(a) of the main text. 
The reference site $j_0$ in the correlation function $C(r) = \braket{\hat{O}^\dagger_{j_0+r(,0)}\hat{O}_{j_0(,0)}}$ is set to $j_0 = L / 4 + 1$ to minimize open boundary effects. 
We use $\hat{O}_{j,\alpha} = \hat{n}_{j,\alpha;\uparrow} - \hat{n}_{j,\alpha; \downarrow}$ for the spin correlation, $\hat{O}_{j,\alpha} = \hat{n}_{j,\alpha} - n_{\rm av}$ for the charge correlation, where $n_{\rm av} = \sum_{j,\alpha} \braket{\hat{n}_{j,\alpha}} / (2L) = 1$, and $\hat{O}_{j,\alpha} = \hat{c}_{j,\alpha;\uparrow} \hat{c}_{j,\alpha; \downarrow}$ for the on-site pairing correlation.
In Fig.~\ref{1}, the results for $m = 8000, 9000$, and $10000$ show almost no difference in the correlation functions, indicating convergence at smaller bond dimensions. 
This rapid convergence can be attributed to the presence of a spin gap, which emerges from spin-singlet formations in the DH pairing phase with a large $t_\perp (= t_\parallel)$.
On the other hand, in Fig.~\ref{0.25}, the on-site pairing and charge correlation functions exhibit a slight increase with increasing $m$, indicating that the calculations have not converged at smaller bond dimensions. 
In contrast to the spin correlations for $t_{\perp}=t_{\parallel}$, which decay exponentially as shown in Fig.~\ref{1}(a), the spin correlation function for $t_{\perp}=0.25t_{\parallel}$ shows a power-law decay, as seen in Fig.~\ref{0.25}(a), indicating that a gapless spin state emerges when $t_{\perp}$ is small. 
Because the spin correlation is dominant compared with the others, we conclude that the small $t_{\perp}$ region in the phase diagram (see Fig.~\ref{tperp}) is the spin density wave (SDW) phase.  
A gapless spin state due to a small $t_\perp (= 0.25t_\parallel)$ may necessitate a large $m$ for sufficient convergence.
Therefore, to ensure numerical accuracy, we consistently use $m=10000$ in the ladder system throughout this study.

\subsection{Averaged correlation functions}

\begin{figure}[b]
  \begin{minipage}[t]{\columnwidth}
      \centering
      \includegraphics[width=\columnwidth]{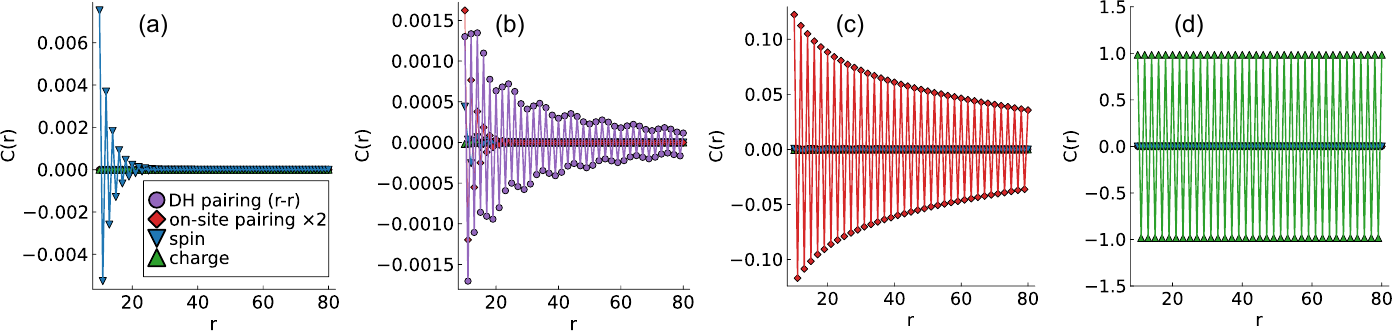}
      \caption{Linear scale plot of the correlation functions using a selected reference site ($j_0=L/4+1$), where $L = 160$, $t_\perp=t_\parallel$, $J_{\perp} = J_{\parallel}=0.4t_\parallel$, and $V = V_{\perp} = V_{\parallel}$. 
      The correlation functions in the (a) spin-singlet phase at $n_d = 0.0, V = 0.4t_\parallel$, (b) DH pairing phase at $n_d = 0.2, V = 0.2t_\parallel$, (c) $\eta$-pairing phase at $n_d = 0.4, V = 0.2t_\parallel$, and (d) CDW phase at $n_d = 0.5, V = 0.8t_\parallel$ are presented.
      }
      \label{rs}
  \end{minipage}
  
  \begin{minipage}[t]{\columnwidth}
      \centering
      \includegraphics[width=\columnwidth]{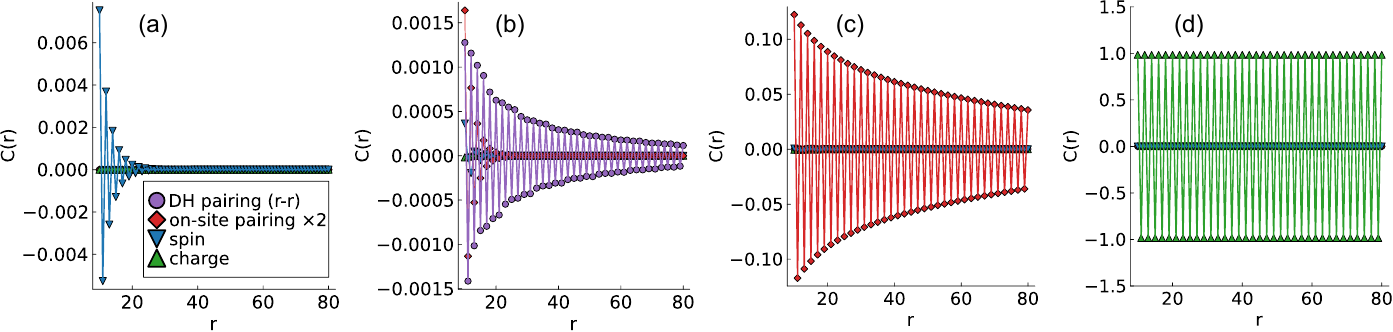}
      \caption{Linear scale plot of the averaged correlation functions.
      All other conditions are the same as in Fig.~\ref{rs}.
      }
      \label{normal}
  \end{minipage}
\end{figure}

The values of the correlation functions calculated under open boundary conditions depend on the choice of the reference site. 
Hence, we present the site-averaged correlation functions.
We define the averaged correlation function for even values of correlation distance $r$ as
\begin{equation}
    \tilde{C}(r) = \frac{1}{8} \sum_{s=-3}^{4} \Braket{\hat{O}^\dagger_{\frac{L-r}{2}+s(,0)}\hat{O}_{\frac{L+r}{2}+s(,0)}},
\end{equation}
and for odd values of $r$ as
\begin{equation}
    \tilde{C}(r) = \frac{1}{9} \sum_{s=-4}^{4} \Braket{\hat{O}^\dagger_{\frac{L-r+1}{2}+s(,0)}\hat{O}_{\frac{L+r+1}{2}+s(,0)}}.
\end{equation}
Figure~\ref{rs} shows the linear scale plot of the correlation functions using a selected reference site ($j_0 = L/4+1$), while Fig.~\ref{normal} shows that of the averaged correlation functions. 
The decay behaviors of the results in Figs.~\ref{rs} and \ref{normal} are almost identical. 
Comparing Figs.~\ref{rs}(b) and \ref{normal}(b), we find that the DH pairing correlation in Fig.~\ref{normal}(b) exhibits a smoother decay, indicating that the decay with a weak oscillation in Fig.~\ref{rs}(b) is due to boundary effects. 
From Figs.~\ref{rs}(a) and \ref{normal}(a), we confirm that the spin correlation decays exponentially in the spin-singlet phase. 
Based on the comparison of the correlation functions at each $(n_d,V)$ point, we constructed the phase diagram in Fig.~1(a) of the main text.

\subsection{$t_{\perp}$ dependence and interchain spin correlation}

\begin{figure}[t]
    \begin{center}       
        \includegraphics[width=0.4\columnwidth]{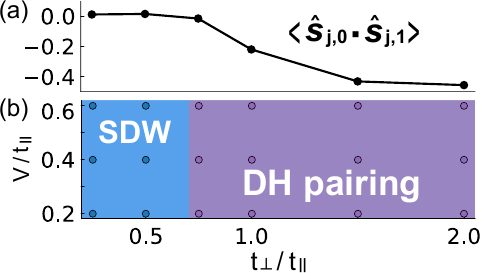}   
        \caption{(a) Interchain spin correlation $\sum_{j} \braket{\hat{\bm{s}}_{j,0} \cdot \hat{\bm{s}}_{j,1}} / L$ for various $t_\perp / t_\parallel$ values, where $V = V_{\perp} = V_{\parallel} = 0.4t_{\parallel}$ and $J_\perp/J_\parallel = (t_{\perp}/t_{\parallel})^2$ with $J_{\parallel}=0.4t_{\parallel}$.
        (b) Phase diagram in the $t_\perp$--$V$ plane. 
        $n_d = 0.1$ and $L = 120$ are used in the calculations.
        }
        \label{tperp}
    \end{center}
\end{figure}

We present supplementary numerical results to examine the role of interchain coupling in the ladder system. 
To understand the role of $t_{\perp}$, we compute the interchain spin correlation $\sum_{j} \braket{\hat{\bm{s}}_{j,0} \cdot \hat{\bm{s}}_{j,1}} / L$ in Fig.~\ref{tperp}(a). 
The interchain spin correlation becomes increasingly negative as $t_\perp/t_\parallel$ increases. 
This reflects the preferable formation of rung spin singlets and the opening of the spin gap as $t_{\perp}/t_{\parallel}$ increases. 
The phase diagram in the $t_\perp$--$V$ plane is shown in Fig.~\ref{tperp}(b). 
As presented in the main text [Fig.~3(a)], the DH pairing correlations increase as $t_\perp/t_\parallel$ increases. 
Hence, we find that the DH pairing becomes dominant at  $t_\perp/t_\parallel \gtrsim 1$.  
On the other hand, when the interchain spin correlation is weak at $t_\perp / t_\parallel < 0.75$, the SDW correlation is dominant as shown in Fig.~\ref{0.25}. 
Hence, a background of rung spin singlets, encouraged by the interchain hopping $t_{\perp}$, is favorable for the development of the DH pairing correlation.   

\section{Derivation of the effective model}
\label{sec:derivation}

\subsection{Preliminaries}

\begin{figure}[b]
    \centering
    \includegraphics[width=0.5\textwidth]{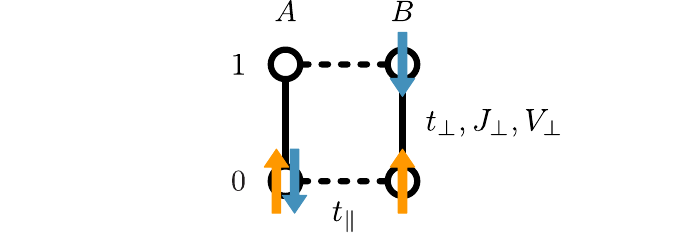}
    \caption{Schematic diagram of the system with a weak interrung hopping $t_{\parallel}$ for $L = 2$, $N_\uparrow = N_\downarrow = 2$, and $N_d = 1$.
    The perturbative $t_\parallel$ term acts along the chain (horizontal) direction, while the nonperturbative $t_\perp$, $J_\perp$, and $V_\perp$ terms act along the rung (vertical) direction.
    }
    \label{min_fig}
\end{figure}

Here, we derive an effective model using perturbation theory within the local rung approximation where $t_{\perp} \gg t_{\parallel}$.
The Hamiltonian with a weak interrung hopping $t_{\parallel}$ is given as follows 
\begin{equation}
  \hat{\mathcal{H}}_{\mathrm{LRA}} = 
  \left[ \hat{\mathcal{H}}^{(0)}_{t_\perp}
  +\hat{\mathcal{H}}_{V_\perp} 
  +\hat{\mathcal{H}}^{(s)}_{J_\perp}
  +\hat{\mathcal{H}}^{(\eta)}_{J_\perp}
  \right]
  +\hat{\mathcal{H}}^{(0)}_{t_\parallel}. 
  \label{min}
\end{equation}
In the main text, we denote 
$\hat{\mathcal{H}}^{(0)}_{t_\perp}
+\hat{\mathcal{H}}_{V_\perp} 
+\hat{\mathcal{H}}^{(s)}_{J_\perp}
+\hat{\mathcal{H}}^{(\eta)}_{J_\perp}$
as $\hat{\mathcal{H}}_{\perp}$ and
$\hat{\mathcal{H}}^{(0)}_{t_\parallel}$ as $\hat{\mathcal{T}}_{\parallel}$. 
We treat the hopping along the chain direction as the perturbation term $\hat{\mathcal{H}}' = \hat{\mathcal{H}}^{(0)}_{t_\parallel}$ and the couplings along the rung direction as the unperturbed term $\hat{\mathcal{H}}_0 = 
\hat{\mathcal{H}}^{(0)}_{t_\perp}
+\hat{\mathcal{H}}_{V_\perp} 
+\hat{\mathcal{H}}^{(s)}_{J_\perp}
+\hat{\mathcal{H}}^{(\eta)}_{J_\perp}$.
For simplicity, we ignore $\hat{\mathcal{H}}_{V_\parallel} + \hat{\mathcal{H}}^{(s)}_{J_\parallel} + \hat{\mathcal{H}}^{(\eta)}_{J_\parallel}$.
Since the perturbation term $\hat{\mathcal{H}}_{t_\parallel}^{(0)}$, which preserves the number of doublons, has no effect on two adjacent rung spin singlets or two adjacent rung DH $\eta$ triplets, we consider the Hilbert space defined by $\hat{\mathcal{H}}_0$ (which we call $\mathcal{K}_0$) with $L = 2$, $N_\uparrow = N_\downarrow = 2$, and doublon number $N_d = 1$.
For $L = 2$, there are two rungs.
We label these two rungs as $A$ and $B$, as shown in Fig.~\ref{min_fig}. 

\subsection{Eigenstates of $\hat{\mathcal{H}}_0$}

Here, we summarize the eigenstates and eigenenergies of $\hat{\mathcal{H}}_0$, which are ingredients for deriving the effective model. 
First, we consider the eigenstates for the particle configurations with $N=1,2,3$ in the single-rung (i.e., two-site) model described by  $\hat{\mathcal{H}}_{\mathrm{rung}}$ in Eq.~(\ref{eq:H_rung}). 
Then, we present the eigenenergy for the two-rung states. 

\subsubsection{$N = 1$, $N_d = 0$}
When $N=1$, there are one singly occupied site and one holon on a rung.
Therefore, we only consider $\hat{\mathcal{H}}^{(0)}_{\mathrm{rung}, t_\perp}$, and the eigenstates and their corresponding energies are as follows
\begin{align}
  \ket{1,1} &= \frac{1}{\sqrt{2}} \left( \hat{c}^\dagger_{0; \uparrow} - \hat{c}^\dagger_{1; \uparrow} \right) \ket{0}, 
  \qquad
  \hat{\mathcal{H}}_{\mathrm{rung}} \ket{1,1} = +t_\perp \ket{1,1},
\\
  \ket{1,2} &= \frac{1}{\sqrt{2}} \left( \hat{c}^\dagger_{0; \downarrow} - \hat{c}^\dagger_{1; \downarrow} \right) \ket{0}, 
  \qquad
  \hat{\mathcal{H}}_{\mathrm{rung}} \ket{1,2} = +t_\perp \ket{1,2},
\\
  \ket{1,3} &= \frac{1}{\sqrt{2}} \left( \hat{c}^\dagger_{0; \uparrow} + \hat{c}^\dagger_{1; \uparrow} \right) \ket{0}, 
  \qquad
  \hat{\mathcal{H}}_{\mathrm{rung}} \ket{1,3} = -t_\perp \ket{1,3},
\\
  \ket{1,4} &= \frac{1}{\sqrt{2}} \left( \hat{c}^\dagger_{0; \downarrow} + \hat{c}^\dagger_{1; \downarrow} \right) \ket{0}, 
  \qquad
  \hat{\mathcal{H}}_{\mathrm{rung}} \ket{1,4} = -t_\perp \ket{1,4}.
\end{align}

\subsubsection{$N = 2$, $N_d = 0$}
When $N = 2$ and $N_d = 0$, there are two singly occupied sites on a rung.
Therefore, we only consider $\hat{\mathcal{H}}^{(s)}_{\mathrm{rung}, J_\perp}$, and the eigenstates and their corresponding energies are as follows
\begin{align}
  \ket{2,1} &= \frac{1}{\sqrt{2}}\left(\hat{c}^\dagger_{0;\uparrow} \hat{c}^\dagger_{1;\downarrow} - \hat{c}^\dagger_{0;\downarrow} \hat{c}^\dagger_{1;\uparrow}\right) \ket{0}, 
  \qquad
  \hat{\mathcal{H}}_{\mathrm{rung}} \ket{2,1} = -J_\perp \ket{2,1},
  \\
  \ket{2,2} &= \frac{1}{\sqrt{2}}\left(\hat{c}^\dagger_{0;\uparrow} \hat{c}^\dagger_{1;\downarrow} + \hat{c}^\dagger_{0;\downarrow} \hat{c}^\dagger_{1;\uparrow}\right) \ket{0}, 
  \qquad
  \hat{\mathcal{H}}_{\mathrm{rung}} \ket{2,2} = 0 \ket{2,2},
  \\
  \ket{2,3} &= \hat{c}^\dagger_{0;\uparrow} \hat{c}^\dagger_{1;\uparrow} \ket{0}, 
  \hspace{92.2pt}
  \hat{\mathcal{H}}_{\mathrm{rung}} \ket{2,3} = 0 \ket{2,3},
  \\
  \ket{2,4} &= \hat{c}^\dagger_{0;\downarrow} \hat{c}^\dagger_{1;\downarrow} \ket{0},  
  \hspace{92.2pt}
  \hat{\mathcal{H}}_{\mathrm{rung}} \ket{2,4} = 0 \ket{2,4}.
\end{align}
Here, $\ket{2, 1}$ corresponds to the previously introduced rung spin-singlet state $\ket{s}$, while $\ket{2,2}$, $\ket{2,3}$, and $\ket{2,4}$ represent the rung spin-triplet states.

\subsubsection{$N = 2, N_d = 1$}
When $N = 2$ and $N_d = 1$, there are one doublon and one holon on a rung.
Therefore, we only consider $\hat{\mathcal{H}}_{\mathrm{rung}, V_\perp} + \hat{\mathcal{H}}^{(\eta)}_{\mathrm{rung}, J_\perp}$, and the eigenstates and their corresponding energies are as follows
\begin{align}
  \ket{2,5} &= \frac{1}{\sqrt{2}}\left(\hat{c}^\dagger_{0;\uparrow} \hat{c}^\dagger_{0;\downarrow} - \hat{c}^\dagger_{1;\uparrow} \hat{c}^\dagger_{1;\downarrow}\right) \ket{0}, 
  \qquad
  \hat{\mathcal{H}}_{\mathrm{rung}} \ket{2,5} = -V_\perp \ket{2,5},
  \\
  \ket{2,6} &= \frac{1}{\sqrt{2}}\left(\hat{c}^\dagger_{0;\uparrow} \hat{c}^\dagger_{0;\downarrow} + \hat{c}^\dagger_{1;\uparrow} \hat{c}^\dagger_{1;\downarrow}\right) \ket{0},  
  \qquad
  \hat{\mathcal{H}}_{\mathrm{rung}} \ket{2,6} = \left(-V_\perp + J_\perp\right) \ket{2,6}.
\end{align}
Here, $\ket{2, 5}$ and $\ket{2, 6}$ correspond to the previously introduced rung DH $\eta$-triplet state $\ket{\eta=1, \eta^z=0}$ and the rung DH $\eta$-singlet state $\ket{\eta=0, \eta^z=0}$, respectively.

\subsubsection{$N = 3, N_d = 1$}
When $N = 3$, there are one singly occupied site and one doublon on a rung.
Therefore, we only consider $\hat{\mathcal{H}}^{(0)}_{\mathrm{rung}, t_\perp}$, and the eigenstates and their corresponding energies are as follows
\begin{align}
  \ket{3,1} &= \frac{1}{\sqrt{2}} \left( \hat{c}^\dagger_{0;\downarrow} \hat{c}^\dagger_{1;\uparrow} \hat{c}^\dagger_{1;\downarrow} - \hat{c}^\dagger_{0;\uparrow} \hat{c}^\dagger_{0;\downarrow} \hat{c}^\dagger_{1;\downarrow} \right) \ket{0}, 
  \qquad
  \hat{\mathcal{H}}_{\mathrm{rung}} \ket{3,1} = -t_\perp \ket{3,1},
  \\
  \ket{3,2} &= \frac{1}{\sqrt{2}} \left( \hat{c}^\dagger_{0;\uparrow} \hat{c}^\dagger_{1;\uparrow} \hat{c}^\dagger_{1;\downarrow} - \hat{c}^\dagger_{0;\uparrow} \hat{c}^\dagger_{0;\downarrow} \hat{c}^\dagger_{1;\uparrow} \right) \ket{0}, 
  \qquad
  \hat{\mathcal{H}}_{\mathrm{rung}} \ket{3,2} = -t_\perp \ket{3,2},
  \\
  \ket{3,3} &= \frac{1}{\sqrt{2}} \left( \hat{c}^\dagger_{0;\downarrow} \hat{c}^\dagger_{1;\uparrow} \hat{c}^\dagger_{1;\downarrow} + \hat{c}^\dagger_{0;\uparrow} \hat{c}^\dagger_{0;\downarrow} \hat{c}^\dagger_{1;\downarrow} \right) \ket{0}, 
  \qquad
  \hat{\mathcal{H}}_{\mathrm{rung}} \ket{3,3} = +t_\perp \ket{3,3},
  \\
  \ket{3,4} &= \frac{1}{\sqrt{2}} \left( \hat{c}^\dagger_{0;\uparrow} \hat{c}^\dagger_{1;\uparrow} \hat{c}^\dagger_{1;\downarrow} + \hat{c}^\dagger_{0;\uparrow} \hat{c}^\dagger_{0;\downarrow} \hat{c}^\dagger_{1;\uparrow} \right) \ket{0},  
  \qquad
  \hat{\mathcal{H}}_{\mathrm{rung}} \ket{3,4} = +t_\perp \ket{3,4}.
\end{align}

\subsubsection{Eigenenergies of the two-rung states}
All two-rung eigenstates of $\hat{\mathcal{H}}_0$ defined in $\mathcal{K}_0$ and their corresponding energies are summarized in Table~\ref{table:eigenstates}.

\begin{table}[htb!]
    \centering
    \footnotesize
    \renewcommand{\arraystretch}{1.2}
    \resizebox{0.8\textwidth}{!}{
    \begin{tabular}{|@{\hspace{2pt}}c@{\hspace{2pt}}|@{\hspace{2pt}}c@{\hspace{2pt}}||@{\hspace{2pt}}c@{\hspace{2pt}}|@{\hspace{2pt}}c@{\hspace{2pt}}||@{\hspace{2pt}}c@{\hspace{2pt}}|@{\hspace{2pt}}c@{\hspace{2pt}}|}
    \hline
    Eigenstate & Eigenenergy & Eigenstate & Eigenenergy & Eigenstate & Eigenenergy \\
    \hline
    $\ket{1, 1}_A \ket{3, 1}_B$ & $0$ & $\ket{2, 1}_A \ket{2, 5}_B$ & $-V_\perp - J_\perp$ & $\ket{3, 1}_A \ket{1, 1}_B$ & $0$ \\
    $\ket{1, 1}_A \ket{3, 3}_B$ & $+2t_\perp$ & $\ket{2, 1}_A \ket{2, 6}_B$ & $-V_\perp$ & $\ket{3, 1}_A \ket{1, 3}_B$ & $-2t_\perp$ \\
    $\ket{1, 2}_A \ket{3, 2}_B$ & $0$ & $\ket{2, 2}_A \ket{2, 5}_B$ & $-V_\perp$ & $\ket{3, 2}_A \ket{1, 2}_B$ & $0$ \\
    $\ket{1, 2}_A \ket{3, 4}_B$ & $+2t_\perp$ & $\ket{2, 2}_A \ket{2, 6}_B$ & $-V_\perp + J_\perp$ & $\ket{3, 2}_A \ket{1, 4}_B$ & $-2t_\perp$ \\
    $\ket{1, 3}_A \ket{3, 1}_B$ & $-2t_\perp$ & $\ket{2, 5}_A \ket{2, 1}_B$ & $-V_\perp - J_\perp$ & $\ket{3, 3}_A \ket{1, 1}_B$ & $+2t_\perp$ \\
    $\ket{1, 3}_A \ket{3, 3}_B$ & $0$ & $\ket{2, 5}_A \ket{2, 2}_B$ & $-V_\perp$ & $\ket{3, 3}_A \ket{1, 3}_B$ & $0$ \\
    $\ket{1, 4}_A \ket{3, 2}_B$ & $-2t_\perp$ & $\ket{2, 6}_A \ket{2, 1}_B$ & $-V_\perp$ & $\ket{3, 4}_A \ket{1, 2}_B$ & $+2t_\perp$ \\
    $\ket{1, 4}_A \ket{3, 4}_B$ & $0$ & $\ket{2, 6}_A \ket{2, 2}_B$ & $-V_\perp + J_\perp$ & $\ket{3, 4}_A \ket{1, 4}_B$ & $0$ \\
    \hline
    \end{tabular}
    }
    \caption{Eigenstates and eigenenergies of $\hat{\mathcal{H}}_0$ defined in $\mathcal{K}_0$.
    }
    \label{table:eigenstates}
\end{table}

\subsection{Effective Heisenberg model}
We assume $V_\perp + J_\perp > |2t_\perp|$; otherwise, as shown in Table \ref{table:eigenstates}, the lowest-energy configuration on a rung would not be a two-fermion state.
Under this condition, the lowest-energy states $\ket{2, 1}_A \ket{2, 5}_B = \ket{s}_A \ket{\eta=1, \eta^z=0}_B$ and $\ket{2, 5}_A \ket{2, 1}_B = \ket{\eta=1, \eta^z=0}_A \ket{s}_B$ are degenerate with $E_0 = -V_\perp - J_\perp$.  
We construct an effective model $\hat{\mathcal{H}}_{\min}$ in the subspace $\mathcal{K}_{\mathrm{min}} (\subset \mathcal{K}_0)$ spanned by $\ket{2, 1}_A \ket{2, 5}_B$ and $\ket{2, 5}_A \ket{2, 1}_B$. 
The projection operator for $\mathcal{K}_{\mathrm{min}}$ is given by
\begin{equation}
  \hat{P} = \ket{2, 1}_{A} \ket{2, 5}_B {}_A\!\bra{2, 1} {}_B\!\bra{2, 5} + \ket{2, 5}_{A} \ket{2, 1}_B {}_A\!\bra{2, 5} {}_B\!\bra{2, 1}. 
\end{equation}
The effective Hamiltonian $\hat{\mathcal{H}}_{\min}$ in the second-order perturbation theory can be derived by
\begin{equation}
  \label{eq:Heffmin}
  \hat{\mathcal{H}}_{\min} = \hat{P} \hat{\mathcal{H}'} \frac{\hat{Q}}{E_{0} - \hat{\mathcal{H}}_0} \hat{\mathcal{H}'} \hat{P} 
  = - t_\parallel^2 \hat{P} \frac{\hat{\mathcal{H}'}}{-t_\parallel} \frac{1}{V_\perp + J_\perp + \hat{\mathcal{H}}_0} \hat{Q} \frac{\hat{\mathcal{H}'}}{-t_\parallel} \hat{P},
\end{equation}
where $\hat{Q} = \hat{1} - \hat{P}$. 
We can calculate Eq.~(\ref{eq:Heffmin}) using the following relations
\begin{align}
  \frac{\hat{\mathcal{H}'}}{-t_\parallel} \hat{P}
  = \frac{1}{2} \Bigl( 
  &\ket{1,1}_A \ket{3,3}_B + \ket{3,3}_A \ket{1,1}_B
  + \ket{1,4}_A \ket{3,2}_B + \ket{3,2}_A \ket{1,4}_B  
  \nonumber \\
-&\ket{1,2}_A \ket{3,4}_B - \ket{3,4}_A \ket{1,2}_B
- \ket{1,3}_A \ket{3,1}_B - \ket{3,1}_A \ket{1,3}_B 
  \Bigr) \times \Bigl( 
  {}_A\!\bra{2,1} {}_B\!\bra{2,5} - {}_A\!\bra{2,5} {}_B\!\bra{2,1} 
  \Bigr),
\end{align}
\begin{align}
  \frac{1}{V_\perp + J_\perp + \hat{\mathcal{H}_0}} \hat{Q} \frac{\hat{\mathcal{H}'}}{-t_\parallel} \hat{P}
= \biggl[ &\frac{1}{2(V_\perp + J_\perp + 2t_\perp)} 
\Bigl( \ket{1,1}_A \ket{3,3}_B + \ket{3,3}_A \ket{1,1}_B 
     - \ket{1,2}_A \ket{3,4}_B - \ket{3,4}_A \ket{1,2}_B \Bigr) 
     \nonumber \\
  + & \frac{1}{2(V_\perp + J_\perp - 2t_\perp)} 
\Bigl( \ket{1,4}_A \ket{3,2}_B + \ket{3,2}_A \ket{1,4}_B 
      - \ket{1,3}_A \ket{3,1}_B - \ket{3,1}_A \ket{1,3}_B \Bigr) 
  \biggr] \nonumber\\
  \times &\Bigl( {}_A\!\bra{2,1} {}_B\!\bra{2,5} 
  - {}_A\!\bra{2,5} {}_B\!\bra{2,1} \Bigr).
\end{align}
Therefore, from Eq.~(\ref{eq:Heffmin}), we obtain
\begin{align}
\hat{\mathcal{H}}_{\mathrm{min}} 
= \left( \frac{-t_\parallel^2}{V_\perp + J_\perp + 2t_\perp} + \frac{-t_\parallel^2}{V_\perp + J_\perp - 2t_\perp} \right) 
\Bigl( & \ket{2,1}_A \ket{2,5}_B {}_A\!\bra{2,1} {}_B\!\bra{2,5} + \ket{2,5}_A \ket{2,1}_B {}_A\!\bra{2,5} {}_B\!\bra{2,1} 
\nonumber \\
- & \ket{2,5}_A \ket{2,1}_B {}_A\!\bra{2,1} {}_B\!\bra{2,5} - \ket{2,1}_A \ket{2,5}_B {}_A\!\bra{2,5} {}_B\!\bra{2,1} 
\Bigr).
\end{align}

\begin{figure}[t]
    \centering
    \includegraphics[width=0.45\textwidth]{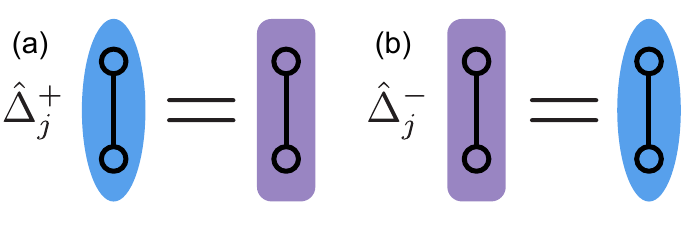}
    \caption{Schematic diagrams of the (a) raising operator $\hat{\Delta}^+_j$ and (b) lowering operator $\hat{\Delta}^-_j$, see Eq.~(\ref{quasi-spin}).
    In comparison with Fig.~\ref{operator}(b), the rung DH pair operator $\hat{\Delta}^{{\rm r}\dag}_j$ corresponds to the raising operator $\hat{\Delta}^{+}_j$.
    }
    \label{quasi-spin_fig}
\end{figure}

Regarding the rung spin-singlet state $\ket{s}=\ket{2,1}$ and the rung DH $\eta$-triplet state $\ket{\eta=1,\eta^z=0}=\ket{2,5}$ as down and up spins, respectively, we define pseudospin operators $\hat{\Delta}^+$, $\hat{\Delta}^-$, and $\hat{\Delta}^z$ as 
\begin{align} \label{quasi-spin_z}
    \hat{\Delta}^z_j \ket{s}_j &= -\frac{1}{2} \ket{s}_j, \quad \hat{\Delta}^z_j \ket{\eta=1, \eta^z=0}_j = +\frac{1}{2} \ket{\eta=1, \eta^z=0}_j,
\end{align}
\begin{align} \label{quasi-spin}
    \hat{\Delta}^+_j \ket{s}_j &= \ket{\eta=1, \eta^z=0}_j, \quad \hat{\Delta}^-_j \ket{\eta=1, \eta^z=0}_j = \ket{s}_j.
\end{align}
The definition of the pseudospin operator $\hat{\Delta}^{+}_j$ ($\hat{\Delta}^{-}_j$) in Eq.~(\ref{quasi-spin}) coincides with the definition of the rung DH pair operators $\hat{\Delta}^{\mathrm{r}\dagger}_j$ ($\hat{\Delta}^\mathrm{r}_j$) in Eq.~(\ref{r-r_DH}), see Fig.~\ref{quasi-spin_fig}.
Using these pseudospin operators, $\hat{\mathcal{H}}_{\mathrm{min}}$ can be written as
\begin{align}
  \hat{\mathcal{H}}_{\mathrm{min}} 
&= \left( \frac{-t_\parallel^2}{V_\perp + J_\perp + 2t_\perp} + \frac{-t_\parallel^2}{V_\perp + J_\perp - 2t_\perp} \right) \left[ \left(\frac{1}{2} - \hat{\Delta}_A^z\right) \left(\frac{1}{2} + \hat{\Delta}_B^z\right) + \left(\frac{1}{2} + \hat{\Delta}_A^z\right) \left(\frac{1}{2} - \hat{\Delta}_B^z\right)
- \hat{\Delta}_A^{+} \hat{\Delta}^{-}_B - \hat{\Delta}^{-}_A \hat{\Delta}_B^{+} \right] \nonumber \\
&= \left( \frac{2t_\parallel^2}{V_\perp + J_\perp + 2t_\perp} + \frac{2t_\parallel^2}{V_\perp + J_\perp - 2t_\perp} \right) \left( \hat{\mathbf \Delta}_A \cdot \hat{\mathbf \Delta}_B - \frac{1}{4} \right).
\end{align}
Under the assumption that $V_\perp + J_\perp > |2t_\perp|$, $\hat{\mathcal{H}}_{\mathrm{min}}$ is equivalent to a two-site antiferromagnetic Heisenberg model.
As mentioned before, rungs are independent in the unperturbed term $\hat{\mathcal{H}}_0$.
Moreover, the perturbation term $\hat{\mathcal{H}}'$ does not affect interactions between two adjacent rung spin singlets or between two adjacent rung DH $\eta$ triplets.
Therefore, for arbitrary chain length $L$ with $N_\uparrow = N_\downarrow = L$ and arbitrary doublon number $N_d$, $\hat{\mathcal{H}}_{\mathrm{min}}$ can be written as
\begin{align} \label{heisenberg}
\hat{\mathcal{H}}_{\mathrm{min}}
= \left( \frac{2t_\parallel^2}{V_\perp + J_\perp + 2t_\perp} + \frac{2t_\parallel^2}{V_\perp + J_\perp - 2t_\perp} \right)
\sum_{j}
\left( \hat{\mathbf \Delta}_j \cdot \hat{\mathbf \Delta}_{j+1} - \frac{1}{4} \right).
\end{align}
Based on the definition of the pseudospin operators in Eq.~(\ref{quasi-spin}), the fixed number of doublons in a photodoped MI corresponds to the fixed pseudomagnetization in Eq.~(\ref{heisenberg}).
Thus, the sign alternation of the DH pairing correlations with a fixed number of doublons can be perturbatively mapped to the sign alternation of the spin correlations $\Braket{\hat{s}^+_{i} \hat{s}^-_{j}}$ in the lowest-energy state of the 1D antiferromagnetic Heisenberg model with a fixed magnetization.

\section{Comparison between Chemically-doped and Photodoped ladders}

We discuss here several key similarities and differences between superconducting pairing in chemically doped systems and DH pairing in photodoped systems.

\begin{figure}[b]
    \centering
    \includegraphics[width=\textwidth]{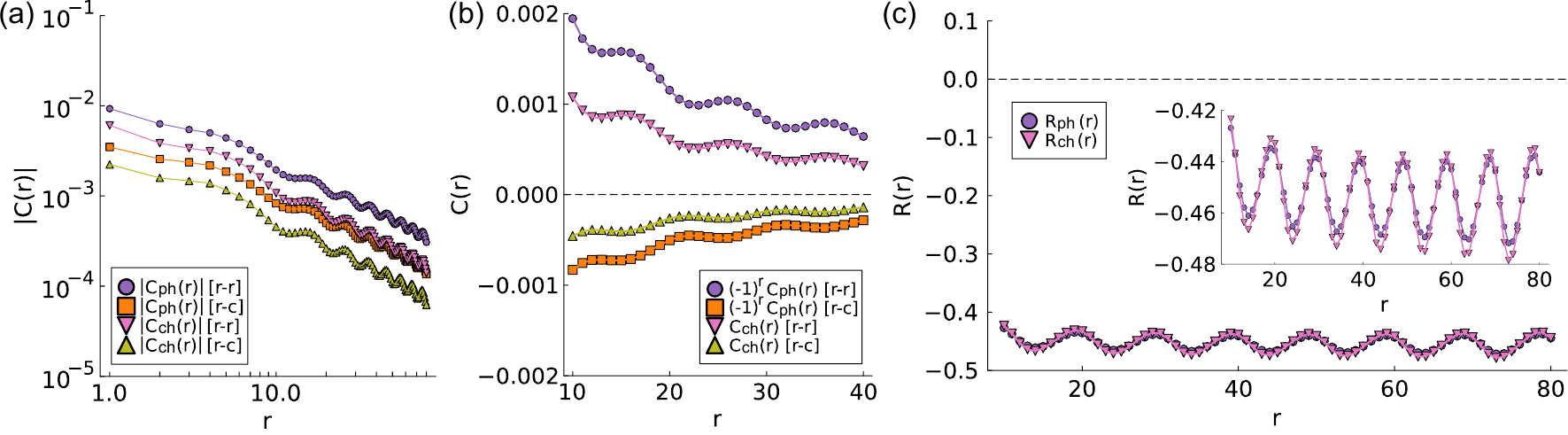}
    \caption{(a) Log-log plot of the absolute values of the pairing correlation functions $C_{\rm ph}(r)$ and $C_{\rm ch}(r)$ for the photodoped and chemically doped ladders, respectively.   
    Purple and orange points represent the rung-rung and rung-chain DH pairing correlations in the photodoped ladder ($n_d=0.05$), while pink and yellow points denote the corresponding HH pairing correlations in the chemically doped ladder (hole density $= 0.1$ per site).   
    The parameters in the model are $t_\perp = t_\parallel$, $J_\perp = J_\parallel = 0.4t_\parallel$, and $V = V_{\perp} = V_{\parallel} = 0.2t_\parallel$.
    (b) Linear scale plot of $(-1)^r C_{\rm ph}(r)$ and $C_{\rm ch}(r)$, where the staggered sign of the DH pairing is removed by multiplying the correlation function $C_{\rm ph}(r)$ by $(-1)^r$. 
    (c) Ratio of rung-chain to rung-rung correlations of the DH pairing state in the photodoped ladder $R_{\rm ph}(r)$ and that of the HH pairing state in the chemically doped ladder $R_{\rm ch}(r)$. 
    Inset: Enlarged view of $R_{\rm ph}(r)$ and $R_{\rm ch}(r)$.
    }
    \label{chem_photo}
\end{figure}

First, we present the similarities. 
Both pairing states arise involving spin fluctuations induced by carrier doping into strongly correlated insulators. 
Particularly, the undoped state in the ladder geometry favors the formation of spin singlets, which can be pairing glues in both chemically doped and photodoped systems.  
Another shared feature is that both pairing states exhibit $d$-wave-like symmetry, where the sign of the rung-chain correlations is opposite to that of the rung-rung correlations. 
To clarify these similarities, we compare the DH pairing correlation in the photodoped case $C_{\rm ph}(r)$ with the HH pairing correlation in the chemically doped case $C_{\rm ch}(r)$. 
As shown in Fig.~\ref{chem_photo}(a), the decays of $|C_{\rm ph}(r)|$ and $|C_{\rm ch}(r)|$ are comparable. 
Since we plot the results with low carrier densities, the spin-singlet formation in the ladder structure should be strongly involved in both pairing states. 
Figure~\ref{chem_photo}(b) compares $(-1)^rC_{\rm ph}(r)$ with $C_{\rm ch}(r)$ in the linear scale plot, where we attach the phase factor $(-1)^r$ to $C_{\rm ph}(r)$ because the DH pairing correlations in the photodoped ladder exhibit sign alternation. 
In both cases, the rung-rung correlations are positive while the rung-chain correlations are negative. 
Figure~\ref{chem_photo}(c) presents the ratio of rung-chain to rung-rung correlations $R(r)=\braket{\hat{\Delta}^{{\rm c} \dag}_{j_0+r} \hat{\Delta}^{{\rm r}}_{j_0}}/\braket{\hat{\Delta}^{{\rm r} \dag}_{j_0+r} \hat{\Delta}^{{\rm r}}_{j_0}}$. 
Corresponding to Fig.~\ref{chem_photo}(b), both $R_{\rm ph}(r)$ for the photodoped case and $R_{\rm ch}(r)$ for the chemically doped case are negative for all values of $r$. 
Notably, $R_{\rm ph}(r)$ and $R_{\rm ch}(r)$ show nearly identical magnitudes. 
This can further support the similarities between DH pairing in the photodoped systems and superconducting pairing in the chemically doped systems. 

Next, we turn to the differences. 
The most crucial difference is that superconducting pairing in chemically doped systems is an equilibrium phenomenon, whereas DH pairing in photodoped systems is a nonequilibrium one. 
While only hole carriers induce superconducting pairing in the chemically doped systems, an equal number of photoinduced doublons and holons contributes to the DH pairing state. 
In this work, we have demonstrated that pairing originating from correlation-driven fluctuations can exist not only in equilibrium systems but also in nonequilibrium photoexcited systems.  
Another key difference is the elements of the pairs.  
The electron-electron or hole-hole pairing correlations exhibit a power-law decay in superconducting pairing, while doublon-holon, i.e., electron-hole, pairing correlations develop in the photodoped condition. 
The doublon-holon pairs possess excitonic characteristics that are charge-neutral and do not carry charge. 
Moreover, while the hole-hole pairing correlations in the chemically doped system are uniform in sign with respect to correlation distance $r$, the DH pairing correlations are staggered. 
As discussed in Sec.~\ref{sec:derivation}, we have explained this behavior by perturbatively connecting the system to the antiferromagnetic Heisenberg model. 
In contrast to the chemically doped system, the $\eta$-spin interaction between doublons and holons activated by photodoping also plays an important role in DH pairing.

\bibliography{Reference}